\documentclass[12pt]{article}
\usepackage{amssymb,amsmath}
\usepackage{epsfig,psfrag}

\catcode`\@=11
\def \tr {\mathop{\rm tr}\nolimits}

\def \e  {\mathop{\rm e}\nolimits}
\newcommand\lr[1]{{\left({#1}\right)}}
\newcommand \widebar [1] {\overline{#1}}

\newcommand \ket [1] {|{#1}\rangle}
\newcommand \bra [1] {\langle {#1}|}
\newcommand\re[1]{(\ref{#1})}
\def \qqquad {\qquad\quad}
\def \qqqquad {\qquad\qquad}

\newcommand{\pa}{\partial}

\newcommand{\be}{\begin{equation}}
\newcommand{\ee}{\end{equation}}
\newcommand{\bea}{\begin{eqnarray}}
\newcommand{\eaa}{\end{eqnarray}}
\newcommand{\nn}{\nonumber}

\renewcommand{\a}{\alpha}

\newcommand{\da}{{\dot\alpha}}
\newcommand{\db}{{\dot\beta}}
\newcommand{\tl}{{\tilde\lambda}}

\renewcommand{\b}{\beta}

\newcommand{\rt}{\tilde\rho}

\newcommand{\la}{\lambda}

\newcommand{\q}{\theta}
\newcommand{\bq}{\bar\theta}

\newcommand{\ep}{\epsilon}

\newcommand{\cN}{{\cal N}}
\newcommand{\cA}{{\cal A}}

\newcommand{\p}[1]{(\ref{#1})}
\newcommand{\bt}[1]{{\bar t}}

\newcommand \vev [1] {\langle{#1}\rangle}
\newcommand \ran [1] {|{#1}\rangle}
\newcommand \lan [1] {\langle{#1}|}

\newcommand{\bi}{{\bar{\imath}}}
\newcommand{\hi}{{\hat{\imath}}}

\newcommand{\hj}{{\hat{\jmath}}}
\newcommand{\bl}{{\tilde\lambda}}
\newcommand{\DT}{{[{\mathcal D}t]_{n,k}}}
\def\numberbysection{\@addtoreset{equation}{section}
                     \def\theequation{\thesection.\arabic{equation}}}
\numberbysection

\begin{document}




\thispagestyle{empty}

\null\vskip-12pt \hfill
\begin{minipage}[t]{30mm}
 IPhT--T10/013 \\
 LAPTH--011/10
\end{minipage}

\vskip3.2truecm
\begin{center}
\vskip 0.2truecm {\Large\bf
Superconformal invariants  for \\[3mm]  scattering amplitudes in   $\cN=4$ SYM theory}

\vskip 1truecm
{\bf  G.P. Korchemsky$^{*}$
 and E. Sokatchev$^{**}$ \\
}

\vskip 0.4truecm
$^{*}$ {\it Institut de Physique Th\'eorique\,\footnote{Unit\'e de Recherche Associ\'ee au CNRS URA 2306},
CEA Saclay, \\
91191 Gif-sur-Yvette Cedex, France\\
\vskip .2truecm $^{**}$ {\it LAPTH\,\footnote[2]{Laboratoire d'Annecy-le-Vieux de Physique Th\'{e}orique, UMR 5108},   Universit\'{e} de Savoie, CNRS, \\
B.P. 110,  F-74941 Annecy-le-Vieux, France
                       }
  } \\
\end{center}

\vskip 1truecm 
\centerline{\bf Abstract} 
\medskip
\noindent
Recent studies of scattering amplitudes in planar $\cN=4$ SYM theory revealed
the existence of a hidden  dual superconformal symmetry. Together with the conventional superconformal symmetry it gives rise to powerful restrictions on the planar scattering amplitudes to all loops. We study the general form of the invariants of both symmetries.
We first construct an integral representation for the most general dual superconformal invariants and show that it allows a considerable freedom in the choice of the integration measure. We then perform a half-Fourier  transform to twistor space, where conventional conformal symmetry is realized locally, derive the resulting conformal Ward identity for the integration measure and show that it admits a unique solution. Thus, the combination of dual and conventional superconformal symmetries, together with invariance under helicity rescalings, completely fixes the form of the invariants. The expressions obtained generalize the known tree and one-loop superconformal invariants and coincide with the recently proposed coefficients of the leading singularities of the scattering amplitudes as contour integrals over Grassmannians.

\newpage

\thispagestyle{empty}

{\small \tableofcontents}

\newpage
\setcounter{page}{1}\setcounter{footnote}{0}

\section{Introduction}\label{Int}

Planar ${\cal N}=4$ super-Yang-Mills theory (SYM) is a remarkable gauge theory with many exceptional properties. It was the first example of an ultraviolet finite four-dimensional field theory possessing (super)conformal symmetry at the
quantum level. Moreover, there is increasing amount of evidence that, in addition
to the ${\cal N}=4$ conformal supersymmetry of the Lagrangian, this theory has
some new symmetries of dynamical origin. This strongly suggests that ${\cal N}=4$ SYM theory is a completely integrable model.

One of the aspects of ${\cal N}=4$ SYM theory which has attracted a lot of attention during the past few years is the study of scattering amplitudes, or equivalently, of
the $S$ matrix elements for a given number of scattered particles. One might think that in a conformal theory of massless fields there is no scale and, therefore, the $S$ matrix should be trivial. Indeed, it is well known that the scattering amplitudes in
a gauge theory suffer from infrared divergences at loop level. When resummed to all loops, the infrared divergences exponentiate in such a way that the scattering
amplitudes vanish after one removes the regularization. In spite of this, there exists
a wealth of different infrared-safe observables (like inclusive cross sections, event shapes, energy-energy correlations, etc) which
are expressed in terms of scattering amplitudes and, at the same time, can be computed order-by-order in the loop expansion. The important difference is that these observables receive contributions from an {\em infinite} number of scattering amplitudes and even though each individual amplitude vanishes due to infrared
divergences, their sum remains finite.

Examining the scattering amplitudes in the dimensionally regularized  ${\cal N}=4$ theory, one discovers a surprisingly rich structure. To start with, the tree (or Born) level amplitudes do not suffer from infrared divergences and, consequently, they exhibit the ${\cal N}=4$ conformal supersymmetry of the Lagrangian. But this is not the whole story. Recently, the study of the simplest MHV and NMHV amplitudes\footnote{The ${\cal N}=4$ supersymmetry Ward identities forbid the existence of scattering amplitudes of $n$ gluons with less than two particles of the same helicity. Thus, the first non-trivial amplitudes involve, say, two gluons of negative and $n-2$ gluons of positive helicity. They are called maximally helicity violating (MHV) amplitudes. More generally, an N${}^k$MHV amplitude involves $k+2$ gluons of negative helicity, with $0 \leq k \leq n-4$. } has revealed the existence of a new hidden symmetry of ${\cal N}=4$ SYM theory, the so-called dual superconformal symmetry \cite{Drummond:2008vq}. {This observation was subsequently extended to all N${}^k$MHV tree-level amplitudes in \cite{Brandhuber:2008pf,Drummond:2008cr}. }
{The strong-coupling counterpart of this new symmetry was identified in \cite{Berkovits:2008ic,Beisert:2008iq} as the so-called fermionic T-duality symmetry of the string sigma model on AdS${}_5\times S^5$.}
Unlike the conventional conformal symmetry, which acts locally on the particle coordinates, the dual conformal symmetry has a local realization on the particle {\it momenta}. When the two symmetries are simultaneously realized in either the coordinate or the momentum representation, one of them is always non-local. This combination of a local with a non-local symmetry gives rise to powerful restrictions on the amplitudes but it is not sufficient to fully fix them.%
\footnote{The closure of the two symmetries has an infinite-dimensional Yangian structure \cite{Drummond:2009fd}, but this does not imply any new constraints on the amplitude.}
Additional assumptions about the analytic properties of the
amplitude are needed \cite{Korchemsky:2009hm,Bargheer:2009qu}.

We wish to point out that dual conformal symmetry manifests itself not only at tree, but also at loop level. In \cite{Drummond:2008vq} it was shown that the so-called NMHV ratio function, obtained by dividing the NMHV six-particle superamplitude by the MHV one, is an exact dual conformal invariant at tree and at one-loop level.   This is a highly non-trivial observation, because the perturbative corrections to the scattering amplitudes suffer from infrared divergences, resulting in a breakdown of conformal symmetry. Then the question is if we can  still extract any useful consequences from the anomalous conformal symmetry at loop level. The non-local nature of the infrared divergences in space-time makes it difficult to analyze the mechanism of breakdown of the local conventional conformal symmetry.

Surprisingly, the {\it non-local} dual conformal symmetry is much better behaved in this aspect. The reason can be traced back to the remarkable duality between gluon scattering amplitudes and lightlike Wilson loops
\footnote{The duality between scattering amplitudes and lightlike Wilson loops was first noticed in QCD in the high-energy (Regge) limit~\cite{Korchemskaya:1996je}.}%
, first proposed by Alday and Maldacena at strong coupling \cite{Alday:2007hr}, and then extended to   perturbation theory in \cite{Drummond:2007aua,Brandhuber:2007yx}. The Wilson loops have a natural conformal symmetry, which turns out to be precisely the dual conformal symmetry of the matching gluon scattering amplitudes. The conformal symmetry of the lightlike Wilson loop is broken by {\it ultraviolet} cusp singularities~\cite{Korchemskaya:1992je}, and its breakdown is much easier to control by means of standard anomalous conformal Ward identities \cite{Drummond:2007cf,Drummond:2007au}. Then the result of  \cite{Drummond:2008vq} is that the dual conformal anomaly has the same universal form for
the MHV and NMHV amplitudes and, therefore, it cancels in the NMHV ratio function. Based on this observation, the conjecture was put forward in  \cite{Drummond:2008vq} that for all N${}^k$MHV superamplitudes in ${\cal N}=4$ SYM, the corresponding ratio function should be dual conformal invariant not only at tree, but even at loop level.

What can we say about the dual {\it supersymmetry} of the ratio
function? This symmetry is present at tree level, but,  quite surprisingly, it is broken already at one loop due to the so-called holomorphic anomaly \cite{Cachazo:2004by,Bena:2004xu}. As was shown in \cite{Korchemsky:2009hm}, the dual supersymmetry of the ratio function is broken in a very
peculiar way. At one loop, the ratio function is given by a sum of
the so-called $R-$invariants, having both conventional and dual superconformal symmetry, multiplied by scalar coefficient functions depending
on the bosonic dual variables only~\cite{Drummond:2008vq,Drummond:2008bq,Brandhuber:2009xz,Elvang:2009ya}.  These functions break the dual superconformal symmetry of the tree-level ratio function down to dual  conformal symmetry. Extending these observations to all loops, we can reduce the
problem of computing the all-loop ratio function to solving two separate problems: finding all invariants of both conventional and dual superconformal symmetries, and then identifying the corresponding coefficient functions. It is the first, and much simpler problem that we address in the present paper.

Recently, Arkani-Hamed {\it et al.} \cite{ArkaniHamed:2009dn} proposed studying a different object, namely the coefficients of the leading singularities of various loop integrals contributing to the scattering amplitudes. Some of these coefficients
already appear at tree and one-loop levels. They coincide with the known
expressions~\cite{Drummond:2008vq,Drummond:2008bq,Drummond:2008cr} for the $R-$invariants mentioned above. Arkani-Hamed {\it et al.}  wrote down a remarkably simple integral formula for the leading singularity coefficients in twistor space, which has manifest conventional superconformal symmetry. Soon afterwards, Mason and Skinner~\cite{Mason:2009qx} came up with a similar integral formula, but this time with manifest dual superconformal symmetry. Neither of these integral representations has both symmetries manifest, but the two formulations were shown to be equivalent in Ref.~\cite{ArkaniHamed:2009vw}.

Still, the question persisted if the new proposal covers all possible invariants of both conventional and dual conformal symmetries, which can appear in the study of ${\cal N}=4$ super-Yang-Mills scattering amplitudes. In the present paper we give an affirmative answer to this question. We start from the formulation of Mason and Skinner~\cite{Mason:2009qx}  and argue that it covers the most general {\em dual} superconformal invariants. This formulation allows a considerable freedom in the choice of the integration measure. We then perform a half-Fourier  transform to twistor space, where conventional conformal symmetry is realized locally. We derive the resulting conformal Ward identity for the integration measure and show that it admits a unique solution, with the specific measure proposed in  \cite{ArkaniHamed:2009dn}. Thus, the combination of dual and conventional superconformal symmetries, together with invariance under helicity rescalings, completely fixes the form of the invariants.

The paper is organized as follows. In Section 2 we review the formulation of
scattering amplitudes in dual superspace. In Section 3 we establish the general
form of the dual superconformal invariants and show that it admits a simple
integral representation in terms of momentum twistors \cite{Hodges:2009hk,Mason:2009qx}, with an
integration measure which is not uniquely fixed.
In Section 4 we perform the twistor (half-Fourier) transform~\cite{Witten:2003nn} of the general dual superconformal invariants and demonstrate that it is localized on a set of intersecting lines in twistor space. In Section 5 we work out the realization of conventional
superconformal symmetry in twistor space, in particular on the auxiliary integration variables. In Section 6 we derive the conformal Ward identity for the integration measure and show that it has a unique solution.  Section 7 contains concluding remarks. Some technical details are summarized in three appendices.

\section{Preliminaries: Scattering amplitudes in dual superspace}

In the on-shell superspace description of scattering amplitudes
in $\mathcal{N}=4$ SYM theory, all asymptotic states (gluons $G^\pm$, gluinos $\Gamma_A, \bar \Gamma^A$ and scalars $S_{AB}$), are combined into a single superstate,
\begin{align} \notag
\Phi(p,\eta) = G^+(p) &+ \eta^A \Gamma_A(p) +\frac12 \eta^A\eta^B S_{AB}(p)
\\ \label{superstate}
& +\frac1{3!} \eta^A\eta^B\eta^C \epsilon_{ABCD} \bar\Gamma^D(p)
+\frac1{4!} \eta^A\eta^B\eta^C\eta^D \epsilon_{ABCD} G^-(p)\,,
\end{align}
with the help of the Grassmann variables  $\eta^A$ carrying an $SU(4)$ index $A=1\ldots4$. The coefficients in the expansion \p{superstate} describe the on-shell states of particles with a lightlike momentum, $p^2=0$, and helicities ranging from $(+1)$ to $(-1)$. Then, all $n-$particle scattering amplitudes can be combined into a single object, the on-shell superamplitude
\be\label{An}
\mathcal{A}_n=\mathcal{A}(\la_1,\bl_1,\eta_1;\ldots;\la_n,\bl_n,\eta_n)\,.
\ee
Here  each scattered superstate is characterized by a pair of commuting two-component spinors $\la_i$ and $\bl_i$,  defining the lightlike momentum
\be\label{p}
p^{\da\a}_i =  \bl^{\da}_i \la^\a_i\,,
\ee
and by the Grassmann variable $\eta_i^A$. The variables $\la,\bl,\eta$ carry helicities $-1/2,1/2,1/2$, respectively.  The expansion of $\mathcal{A}_n$ in powers of the $\eta$'s generates scattering amplitudes for the various types of particles.
The $SU(4)$ invariance of $\mathcal{A}_n$ combined with the on-shell Poincar\'e supersymmetry imply that the expansion has the following
form
\begin{align}\label{An-dec}
\mathcal{A}_n = \mathcal{A}_n^{\rm MHV} +  \mathcal{A}_n^{\rm NMHV}
+\ldots+\mathcal{A}_n^{\rm \widebar{MHV}}\,,
\end{align}
where each term is a homogenous polynomial in the $\eta_i^A$ of degree
$8+4k$ with $k=0,\ldots,n-4$ {referring to $\mathcal{A}_n^{\rm N^kMHV} $}. The first term in the expansion, $\mathcal{A}_n^{\rm MHV}$,  is of degree 8 and generates all $n-$particle MHV amplitudes. The next term of the expansion, $\mathcal{A}_n^{\rm NMHV}$, has degree $12$ in $\eta$ and generates the NMHV amplitudes, etc.

As an example, consider the simplest of all superamplitudes, the tree-level MHV one, in the form proposed by Nair \cite{Nair:1988bq}:
\begin{align}\label{mhv}
\cA_n^{\rm MHV;0} =  \frac{\delta^{(4)}(\sum_1^n \la_i\bl_i) \delta^{(8)}(\sum_1^n \lambda_i\eta_i) }{\vev{12}\vev{23}\ldots \vev{n1}}\,.
\end{align}
In the numerator we find the delta functions of (super)momentum conservation. The denominator contains Lorentz invariant contractions of spinors $\la_i$ (see Appendix A for the notation), which gives the superamplitude the necessary helicity weight $(+1)$ at each point. Expanding the Grassmann delta function, we get different component amplitudes, whose helicity structure is determined by the combination of variables $\eta_i$ in each term of Grassmann degree 8.

As mentioned in the introduction, the tree-level scattering amplitudes  inherit the superconformal symmetry of the $\mathcal{N}=4$ SYM Lagrangian.  In momentum superspace this symmetry is realized
{\em non-locally} on the variables $(\la, \tilde\la, \eta)$, with generators in the form of second-order differential operators   \cite{Witten:2003nn}. In addition to this conventional superconformal symmetry, the {planar} scattering amplitudes have another, {\it dual $\cN=4$ superconformal symmetry}. To exhibit this symmetry one introduces new dual variables \cite{Broadhurst,Drummond:2006rz,Drummond:2008vq}
 related to the supermomenta
$(p_i,\eta_i)$ as follows
\begin{align}
& p^{\da\a}_i
 = x^{\da\a}_i - x^{\da\a}_{i+1}\,, \qqqquad
\la_i^\a \eta_i^A = \theta_i^{\a A}-\theta_{i+1}^{\a A}\,,
\label{super-mom}
\end{align}
with the periodicity conditions $x_{n+1} \equiv x_1$  and $\theta_{n+1}  \equiv \theta_{1}$.
The dual superconformal symmetry acts {\it locally} on the dual variables $x_i$ and $\theta_i$  as if they were coordinates in some dual superspace. Most remarkably, the tree-level superamplitude \p{An-dec}, rewritten in terms of the dual coordinates, transforms covariantly under dual superconformal symmetry with dual conformal weight $(+1)$ at each point, equal to the helicity of the superstate \re{superstate}. At loop level, the dual superconformal invariance is broken by quantum corrections and the corresponding anomalies have been studied in Refs.~\cite{Drummond:2007au,Brandhuber:2009xz}.

To make the conventional and dual superconformal symmetries manifest, it is convenient to rewrite \re{An-dec} in the equivalent factorized form
\begin{align}\label{R}
\mathcal{A}_{n}  = {\cA}_{n}^{\rm MHV}\left[1+ R^{\rm NMHV} + \ldots + R^{{\rm N}^k {\rm MHV}}+ \ldots +  R^{\rm \widebar{MHV}}\right],
\end{align}
where $R^{{\rm N}^k {\rm MHV}}$ with $k=0,1,\ldots, n-4$ are the so-called  `ratio functions' of Grassmann degree $4k$. The reason for introducing the ratio functions is the following.
The total {planar} superamplitude $\mathcal{A}_{n}$ and its MHV component
${\cA}_{n}^{\rm MHV}$ have the same infrared divergences, carry the helicities of the $n$ scattered particles and their dual conformal weights. As a consequence, the ratio functions are free from infrared divergences to all loops, they have vanishing helicity and, most importantly, they are invariant under both conventional and dual superconformal transformations at tree level.

The first superconformal invariants of this type were discovered  in  \cite{Drummond:2008vq} by inspecting the properties of the NMHV tree superamplitudes formulated in dual superspace:
\begin{align}\label{mbubl}
    R_{cab} = \frac{\vev{a-1\, a} \vev{b-1\, b}\ \delta^{(4)}\left(\lan{c} x_{cb} x_{ba}\ran{\q_{ac}}  +  \lan{c} x_{ca} x_{ab}\ran{\q_{bc}} \right)}{x^2_{ab}\lan{c} x_{cb} x_{ba}\ran{a-1}\lan{c} x_{cb} x_{ba}\ran{a} \lan{c} x_{ca} x_{ab}\ran{b-1}\lan{c} x_{ca} x_{ab}\ran{b}}\,,
\end{align}
where the indices $a-1, a, b-1, b$ and $c$ label five external particles. The expression \re{mbubl} can be thought of as the supersymmetric extension of the three-mass box coefficients found in Ref.~\cite{Bern:2004bt}. It is a homogenous polynomial in the $\theta$'s of degree 4 taking the special form of a Grassmann delta function. It terms of the
invariants \re{mbubl}, the tree-level NMHV superamplitude takes a remarkably simple form,
\begin{align}\label{R-NMHV}
R^{\rm NMHV;0} = \sum_{4 \leq a+1 < b \leq n} R_{1ab}
\,.
\end{align}
The $R-$invariants \re{mbubl} satisfy nontrivial relations~\cite{Drummond:2008vq,Drummond:2008bq,Brandhuber:2009xz} which ensure the invariance of \re{R-NMHV} under a cyclic shift of the labels of the particles, $i\mapsto i+1$.

In a subsequent development, the complete tree-level superamplitude \p{R} was derived in Ref.~\cite{Drummond:2008cr} from the supersymmetric version of the BCFW recursion relations~\cite{Britto:2005fq,Bianchi:2008pu,Brandhuber:2008pf,ArkaniHamed:2008gz,Elvang:2008na}. It is expressed in terms of dual superconformal invariants of Grassmann degree $4k$, similar in structure to the simplest $k=1$ invariant \p{mbubl}.

Mason and Skinner \cite{Mason:2009qx} (inspired by Arkani-Hamed {\it et al.} \cite{ArkaniHamed:2009dn} and Hodges \cite{Hodges:2009hk}) proposed an elegant reformulation of the invariant \p{mbubl}, suitable for immediate generalization to any N${}^k$MHV amplitudes. They considered the integral
\begin{equation}\label{msin}
   R^k_n(W) = \int \DT \prod_{a=1}^k\delta^{(4|4)}\bigg(\sum_{i=1}^n t^i_a W_i\bigg)\,,\qquad
   W_i = \left(\begin{array}{r} \la_{i} \\
    x_i \la_{i} \\
    \q_{i} \la_{i}
    \end{array} \right) \,.
\end{equation}
Here $W_i$
are the so-called momentum supertwistors \cite{Hodges:2009hk} which transform homogeneously under the linear action of the superconformal group $SL(4|4)$ (see Sect.~\ref{gdsci} for more detail). The integration variables $t_a^i$  form a $k\times n$ matrix of {\it real} \footnote{In the presentation of Ref.~\cite{Mason:2009qx}, as well as in Ref.~\cite{ArkaniHamed:2009dn}, the $t$'s are taken to be complex variables, and the integral goes along a contour specific for each type of invariant. In this paper we shall carry out a twistor (Fourier) transform of the variables $\tilde\la$, therefore we need to deal with real variables and real delta functions. For the same reason, we work in space-time with split signature $(++--)$ and the superconformal group is $SL(4|4)$, instead of the more familiar $SU(2,2|4)$ in Minkowski space-time.  The results we obtain can be compared to those of other approaches by analytic continuation. }  integration variables (the indices $a=1\ldots k$ and $i=1\ldots n$ transform under $GL(k)$ and $GL(n)$, respectively).

The basic idea behind this proposal is very simple: The $k$ linear combinations $\sum_{i=1}^n t^i_a W_i$ of supertwistors are simply rotated by $SL(4|4)$ supermatrices, so the invariance of \p{msin} is manifest (the coefficients of the linear combinations are supposed inert). The role of the integral $\int \DT$ over the $t$'s (with a measure $\DT$ to be discussed below) is to make $R^k_n$ {be} a function of the (super)momentum variables $(\la_i, \tilde\la_i, \eta_i)$ only. How to actually carry out the integration requires further discussion. In very simple cases (e.g., for $k=1, n=5$), when the number of bosonic delta functions in  \p{msin} matches the number of integrations, the integral can be done directly. In the generic case the integral is treated as a complex one with a specifically chosen contour for each type of superinvariant to be produced. We refer the reader to \cite{ArkaniHamed:2009dn,Mason:2009qx} for the details. For our purposes here we need not evaluate any of these integrals.

In the following section we will argue that the integral representation of the type \p{msin} describes the most general dual superconformal invariants of Grassmann degree $4k$. The only freedom in them is confined to the measure $\DT$, which we have not specified yet. In the rest of the paper we will show that the requirement of conventional conformal invariance uniquely fixes this measure.

\section{The general dual superconformal invariant}\label{gdsci}

The problem of finding superconformal invariants for scattering amplitudes
in $\cN=4$ SYM theory can be formulated as follows. We are looking
for functions of the supermomenta $(\lambda_i,\tilde\lambda_i ,\eta_i)$ of $n$ particles, which satisfy three conditions of invariance under:
\begin{itemize}
\item helicity rescalings
\begin{align}\label{hel1}
  \la_i \to \zeta_i\la_i\,, \qquad \tilde\lambda_i \to \zeta_i^{-1}\tilde\la_i\,, \qquad \eta_i \to \zeta_i^{-1}\eta_i\,;
\end{align}
\item conventional $SL(4|4)$ superconformal transformations;
\item dual $SL(4|4)$ superconformal transformations.
\end{itemize}
One of the difficulties in implementing these conditions is due to the fact
that the generators of  conventional superconformal transformations
act on $(\lambda_i, \tilde\lambda_i,\eta_i)$ as second-order differential operators and, as a consequence, the corresponding transformations
are non-local. This is in contrast to the {\em linear} dual superconformal transformations whose generators are first-order differential operators
acting on the dual variables.
The conventional superconformal transformations become local after performing
a twistor (half-Fourier) transform~\cite{Witten:2003nn}. This does not make the problem simpler, however, since then the dual superconformal transformations become non-local upon the
twistor transform.

We will implement the above mentioned conditions in two steps. We will first construct helicity neutral functions of $(\lambda_i,\tilde\lambda_i ,\eta_i)$ invariant
under   dual superconformal transformations. As we will see, there is considerable freedom in the choice of these functions. Then, we will impose the condition of conventional superconformal invariance and will show that it removes the ambiguity. In this way, we will arrive
at the general expression for the superconformal invariants.

\subsection{Dual superconformal symmetry}\label{dssms}

The dual superconformal symmetry of scattering amplitude was discovered in  \cite{Drummond:2008vq} by inspecting the properties of MHV and NMHV tree superamplitudes formulated in dual superspace. This symmetry acts locally on the  {\it dual superspace coordinates} introduced in \p{super-mom}. Putting together \re{p} and \re{super-mom}, we express the dual variables in terms of $(\lambda_i,\tilde\lambda_i ,\eta_i)$ as follows:
\begin{eqnarray}
  x_{i,i+1} =  \tl_i \la_i &\rightarrow& x_i = x_1 - \sum_{j=1}^{i-1} \tl_j \la_j \,, \nn \\
  \q^A_{i,i+1} = \eta^{i\, A} \la_i  &\rightarrow& \q^A_i = \q^A_1 - \sum_{j=1}^{i-1}  \eta_j^{A}  \la_j \label{2} \ ,
\end{eqnarray}
with some arbitrary $x_1$ and $\q_1$ (dual superspace translation invariance). We recall that $\tilde\lambda_i$ and $\eta_i^A$ carry   helicity $1/2$ whereas $\lambda_i$ has helicity $-1/2$. Then it follows from \re{2}
that $x_i$ and $\theta_i^A$ have vanishing helicity.

Let us briefly recall how dual superconformal symmetry acts in this superspace.
The dual Poincar\'e supersymmetry is realized in the standard (chiral) form
\begin{equation}\label{posusy}
    Q_{A\, \a} \q_i^{B\, \b} = \delta^B_A \delta^\b_\a\,, \qqquad \bar Q^A_{\da} x_i^{\db\b} = i\delta^{\db}_{\da} \q_i^{A\, \b}\,, \qqquad P_{\a\da} x_i^{\db\b} = \delta^\b_\a\delta^{\db}_{\da} \,.
\end{equation}
As an example, let us consider the NMHV invariant \re{mbubl}.
It is straightforward to verify that it is invariant under $Q$ and $P$, but showing  invariance under $\bar Q$ is less trivial \cite{Drummond:2008vq}. The essential point is the presence of the Grassmann delta function in the numerator of \p{mbubl}, which suppresses the $\bar Q$ variation of the denominator (see a more detailed discussion in Sect.~\ref{tgcmsi} below).

Dual (super)conformal symmetry can be regarded as the dual (super)Poincar\'e group enhanced by the discrete operation of conformal inversion $I$, satisfying the relation $I^2=\mathbb{I}$. The proper dual conformal transformations (boosts) are obtained by combining inversion and translation, $K=IPI$ (and similarly for the dual superconformal generators,  $S=I\bar Q I$ and $\bar S = IQI$). Thus,
in order to prove the full dual superconformal invariance of the function  \re{mbubl}, which is annihilated by the super-Poincar\'e generators \p{posusy}, it is sufficient to show its invariance under inversion.
The action of inversion on the dual coordinates was formulated
in \cite{Drummond:2008vq}:
\begin{equation}\label{duinv}
    I: \qquad (x_i^{\da\a})' = (x_i^{-1})_{\a\da}\,, \qquad (\la_{i}^\a)' = x^{\da \b} \la_{i\b} \,, \qquad (\q_i^\a)' = \q_{i}^\b (x_i^{-1})_{\b\da}\,.
\end{equation}
It is then easy to check that the expression \p{mbubl} is indeed
invariant under inversion \re{duinv}.

Dual (super)conformal symmetry is made more transparent by introducing the notion of {\it momentum} supertwistors  \cite{Hodges:2009hk,Mason:2009qx}
\begin{equation}\label{ho}
  W_i = \left(\begin{array}{l} w^{\hat A}_i  \\
    \chi^A_i
    \end{array} \right) \,,\qquad
    w^{\hat A}_i = \left(\begin{array}{l} \la^\a_i \\
    \nu_{i\,\da}
    \end{array} \right) \equiv \left(\begin{array}{r} \ket{i} \\
    x_i\ket{i}
    \end{array} \right)
    \,,\qquad \chi^A_i=\q^A_{i}\ket{i}\,.
\end{equation}
Here $w^{\hat A}$ and $\chi^A$ are the bosonic and fermionic components of
the supertwistor, respectively. They both carry  indices $\hat A, A=1,\ldots,4$ in the fundamental representation of $SL(4)$, but their meaning is different.
For $w^{\hat A}$, the corresponding $SL(4)$ is the conformal symmetry  group in a space-time with split signature $(++--)$, while for $\chi^A$ it is the R-symmetry group of ${\cal N}=4$ conformal supersymmetry.
The dual coordinates are expressed in terms of the components of the momentum supertwistors
as follows,
\begin{align}\label{tr}
x_i =\frac{|\nu_{i-1}]\, \bra{i}-|\nu_{i}]\, \bra{i-1}}{\vev{i\,i-1}}\,,\qqqquad \theta_i^A = \frac{ \chi_{i-1}^A \bra{i}- \chi_{i}^A \bra{i-1}}{\vev{i\,i-1}}\,,
\end{align}
where we have used the conventions from Appendix~A. The transformation properties of the components of the supertwistor  $W_i$  under conformal inversion  follow directly from \p{duinv}: \footnote{We remark that there exists an ambiguity in ascribing a conformal weight to $\la$ \cite{Drummond:2008vq}. For instance, in \p{duinv} we could choose $\la'_i = x^{-1}_i\ket{i}$, resulting in the definitions $\nu_i = x^{-1}_i\ket{i}$ and $\chi_i = x^{-2}_i \q_i\ket{i}$, without affecting the supertwistor transformation rule \p{18ho}. This alternative choice would give more natural dilatation weights to $\nu$ and $\chi$, but it would considerably complicate the expression \p{1} of the amplitude in momentum space. As because we plan to Fourier transform \p{1} with respect to $\tilde\la$, we prefer to keep to the simplest definitions \p{duinv}.}
\begin{equation}\label{18ho}
    I: \qquad (\la_i^\a)' = \nu_{i\,\da}\,, \qquad (\nu_{i\,\da})' = \la_i^\a \,, \qquad ({\chi^A_i})' = \chi^A_i\,.
\end{equation}
It is easy to see that this transformation has the basic property of inversion $I^2= \mathbb{I}$.

The advantage of the momentum supertwistor notation is that the action of the dual superconformal algebra consists of {\it linear} transformations of $W_i$ with generators
\begin{align}
  & {\mathcal Q}^{\hat A}_A = \sum_{i=1}^n w_i^{\hat A} \frac{\pa}{\pa\chi_i^A}\,, &&
  \hspace*{-20mm}
  \tilde {\mathcal Q}^{A}_{\hat A} = \sum_{i=1}^n \chi_i^A \frac{\pa}{\pa w_i^{\hat A}}\,,  \nn \\
  &  {\mathcal M}^{\hat A}_{\hat B} = \sum_{i=1}^n w_i^{\{\hat A} \frac{\pa}{\pa w_i^{\hat B\}}}\,, &&\hspace*{-20mm}  {\mathcal N}^A_B = \sum_{i=1}^n \chi_i^{\{ A} \frac{\pa}{\pa\chi_i^{B\}}}\,, \nn\\
  &  {\mathcal C} =  \sum_{i=1}^n\left(\sum_{\hat C=1}^4 w_i^{\hat C} \frac{\pa}{\pa w_i^{\hat C}}   + \sum_{ C=1}^4 \chi_i^{ C} \frac{\pa}{\pa \chi_i^{ C}} \right) \,, \label{ducsusygen}
\end{align}
where $\{\}$ denotes the traceless part. Here ${\mathcal Q}=(Q,\bar S)$ and $\tilde {\mathcal Q}=(\bar Q, S)$ are odd (superconformal) generators, ${\mathcal M}$ are the generators of the dual conformal $SL(4)$ transformations,
${\mathcal N}$ are the generators of R-symmetry and
 ${\mathcal C}$ is the central charge generator in the ${\cal N}=4$ dual superconformal algebra  $SL(4|4)$,
\begin{equation}\label{ccal}
    \{ {\mathcal Q}^{\hat A}_A, \tilde {\mathcal Q}^{B}_{\hat B}\}= \delta^B_A \,{\mathcal M}^{\hat A}_{\hat B} + \delta^{\hat A}_{\hat B}\, {\mathcal N}^B_A + \frac1{4} \delta^B_A \delta^{\hat A}_{\hat B}\, {\mathcal C}\,.
\end{equation}
The central charge ${\mathcal C}$ can be identified with the total helicity of the amplitude \cite{Drummond:2008vq}.

In summary, the superinvariants $R^k_n$ that we are looking for should be functions of the momentum supertwistors $W_i$, invariant under global (point-independent) $SL(4|4)$ rotations. In addition,  $R^k_n$ should be invariant
under local helicity transformations, i.e. individual rescalings of each momentum supertwistor $W_i$, corresponding to the helicity transformations \re{hel1},
\begin{align}\label{W-h}
W_i \to \zeta_i\, W_i\,.
\end{align}
We remark that this local helicity invariance automatically implies invariance
of $R^k_n$ under the global (point-independent) transformations $W_i \to \zeta \, W_i $
generated by the central charge ${\mathcal C}$. In the next subsection we will find the general form of such invariants.

\subsection{Chiral dual superconformal invariants}

Before we embark on the construction of the general dual superconformal invariants, let us
look more closely at the NMHV invariant \p{mbubl}. By construction, it must be a homogenous polynomial in the Grassmann variables of degree 4.
A characteristic feature of $R_{cab}$ is that this polynomial has the special form of a Grassmann delta function. This is not accidental --  as was pointed out in Ref.~\cite{Drummond:2008vq}, the chiral invariants of Poincar\'e supersymmetry should necessarily involve Grassmann delta functions.

To elucidate the reason for this, let us go back to \p{mbubl} and reexpress $R_{cab}$ in terms of momentum supertwistors \p{ho} with the help of \p{tr}. {The resulting expression is (see Ref.~\cite{Mason:2009qx})
\begin{equation}\label{rtwi}
    R_{cab}= \frac{\delta^{(4)}(\chi_c \vev{a-1,a,b-1,b} + {\rm cycle})}{\vev{a-1,a,b-1,b}\vev{a,b-1,b,c}\vev{b-1,b,c,a-1}\vev{b,c,a-1,a}\vev{c,a-1,a,b-1}}\,,
\end{equation}
with $\vev{a,b,c,d} = \ep_{{\hat A}{\hat B}{\hat C}{\hat D}} w^{\hat A}_a w^{\hat B}_b w^{\hat C}_c w^{\hat D}_d $. It
depends on five supertwistors $W_{a-1}$,
$W_a$, $W_{b-1}$, $W_b$ and $W_c$, so {that} the dual superconformal generators \p{ducsusygen} act on these five points only.} It is straightforward to verify the invariance of $R_{cab}$ under the generators ${\mathcal Q}^{\hat A}_A$ defined in \p{ducsusygen}. As in  Ref.~\cite{Drummond:2008vq},  we then make use of the 16 odd parameters of the ${\mathcal Q}-$symmetry  to shift away  the Grassmann components of four out of the five $W$'s:
\begin{align}\label{gauge}
\chi_{a-1}^A=\chi_{a}^A=\chi_{b-1}^A=\chi_{b}^A=0\,.
\end{align}
Substituting these relations into \p{rtwi}, we find that $R_{cab}\sim \delta^{(4)}(\chi_c)$.
In the frame \p{gauge}, the generators $\tilde {\mathcal Q}^{A}_{\hat A}$, Eq.\,\p{ducsusygen}, only act on  $w_c$, with variations proportional to $\chi_c^A$. However, $\chi_c^A$ is annihilated by the Grassmann delta function $\delta^{(4)}(\chi_c)$, so that $R_{cab}$ stays invariant under the $\tilde {\mathcal Q}^{A}_{\hat A}$ transformations. As we will see soon, the same mechanism is at work for the most general dual superconformal invariants. The basic reason for having Grassmann delta functions is the chiral realization of supersymmetry.

\subsubsection{Analogy with propagators of chiral and antichiral superfields}\label{toy}

As a simple and well-known illustration how to build invariants of Poincar\'e supersymmetry,  consider the propagators (two-point functions) of (anti)chiral matter superfields
in an $\cN=1$ (massive) Wess-Zumino model, $\Phi(z,\q)$, and its conjugate  $\bar\Phi(\bar z,\bq)$. Here $(z = x + i\q\bq, \ \q)$ and $(\bar z = x - i\q\bq, \ \bq)$ form the so-called chiral and antichiral bases in superspace, respectively, which are closed under the action of  Poincar\'e supersymmetry:
\begin{align}  \label{55'}
  &    \delta z^{\a\da} = 2i\q^\a \bar\ep^{\da}\,, \qqqquad  \ \ \,  \delta\q^\a = \ep^\a  \,,
  \\[2mm]
  &   \delta \bar z^{\a\da} = -2i\ep^\a \bq^{\da}\,, \qqqquad
    \delta\bq^{\da} = \bar\ep^{\da} \,.   \label{55''}
\end{align}
The manifestly supersymmetric propagators of these superfields are
\begin{eqnarray} \label{54}
  \vev{\Phi(z_{1},\q_1) \bar\Phi(\bar z_{2},\bq_2)} &\propto& \frac{1}{(z_{1}-\bar z_{2} - 2i\q_1\bq_2)^2} \,,
  \\   \label{55}
  \vev{\Phi(z_{1},\q_1) \Phi(z_{2},\q_2)} &\propto& \frac{  \delta^{(2)}(\q_1-\q_2)}{(z_{1}-z_{2} )^2}\,,
\end{eqnarray}
and the complex conjugate of the second relation.
In the presence of both chiral and antichiral superfields it is possible to construct a supersymmetric invariant interval (the denominator in \p{54}). If one uses only chiral (or only antichiral) superfields, such an interval does not exist
and the Grassmann coordinates {can only} enter through a delta function.

{To explain this phenomenon, it is helpful to fix appropriate frames in superspace, in which the supersymmetry generators are ``frozen". In the non-chiral realization \p{55'} and \p{55''} we can use both supersymmetry parameters $\ep$ and $\bar\ep$ to shift away both Grassmann coordinates,  i.e. to set $\q_1=0$ and $\bq_2=0$. In this frame $\delta z_1=\delta \bar z_2=0$, but we still have to impose translation invariance. Thus, the supersymmetric and translation invariant we have constructed is $z_1-\bar z_2$.  Undoing the frame fixing, i.e. performing the inverse supersymmetry transformation with parameters which restore $\q_1$ and $\bq_2$, we obtain precisely the supersymmetric interval $z_{1}-\bar z_{2} - 2i\q_1\bq_2$  in \p{54}. In the purely chiral realization \p{55'} we can use the parameter $\ep$ to shift away only one of the $\q$'s, e.g. $\q_1=0$. In this frame $\delta z_1=0$, but the remaining $z_2$ transforms,  $\delta z_{2} = 2i  \q_2 \bar\ep \neq 0$ with an arbitrary $\bar\ep$, and we have nothing to compensate this variation with. So, in chiral superspace
the only way to construct a supersymmetric invariant two-point function is to include a Grassmann delta function, $\delta^{(2)}(\q_2)$, which suppresses $\delta z_{2}$. Undoing the frame $\q_1=0$, we obtain the numerator in \p{55}. Of course, Poincar\'e supersymmetry alone does not fix the dependence on the bosonic variable $z_1-z_2$ (the denominator in \p{55}).}

Although the two supersymmetric invariants \p{54} and \p{55} seem quite different, there exists an integral transform which relates the former to the latter. Consider the antichiral superspace integral
\begin{equation}\label{intr}
    \int d^4\bar z_3 d^2\bq_3 \ \vev{\Phi(z_{1},\q_1) \bar\Phi(\bar z_{3},\bq_3)}\vev{\bar\Phi(\bar z_{3},\bq_3) \Phi(z_{2},\q_2) }\,.
\end{equation}
The measure is clearly invariant under \p{55''}, so we can expect the result to be a supersymmetric chiral two-point function.  Indeed, the integration can be easily performed in the gauge $\q_1=0$,
\begin{equation}\label{intr'}
    \int  \frac{d^4\bar z_3 d^2\bq_3}{(z_1-\bar z_3)^2} \ e^{2i\q_2 \bq_3 \pa_{\bar z_3}} \frac1{ (\bar z_3 - z_2)^2} \propto  \int  \frac{d^4\bar z_3 }{(z_1-\bar z_3)^2} \ (\q_2)^2 \Box_3 \frac1{ (\bar z_3 - z_2)^2}  \propto \frac{  \delta^{(2)}(\q_2)}{(z_1-z_{2} )^2}\,,
\end{equation}
and we recover \p{55}.

\subsubsection{ Chiral (holomorphic) supertwistor invariants}\label{hsti}

The situation in momentum supertwistor space is very similar. In a close analogy with the
chiral and antichiral superfields, we can consider two kinds of supertwistors. The first one is a holomorphic (or chiral) supertwistor
\begin{equation}\label{hotw}
    W =
\left(\begin{array}{c}\la^\a \\\nu_{\da} \\\chi^A\end{array}\right)  =
 \left(\begin{array}{c}w^{\hat A} \\\chi^A\end{array}\right) \,,
\end{equation}
and the second one is antiholomorphic,
\begin{equation}\label{hotw'}
    \widebar W=(\bar\lambda_\alpha,\bar\nu^\da,\bar\chi_A) \equiv(\bar w_{\hat A}, \bar\chi_A)\,,
\end{equation}
belonging to the conjugate $SL(4|4)$ representation. Having these two types of supertwistors, we can easily construct dual superconformal $SL(4|4)$ invariants  in the form of an inner product,
\begin{align}
W\cdot \widebar W \equiv w^{\hat A} \bar w_{\hat A}+\chi^A\bar\chi_A\,.
\end{align}
We may say that this is the analog of the chiral-antichiral two-point function \p{54}. We recall, however, that our description of ${\cal N}=4$ superamplitudes is purely holomorphic (see \re{superstate} and \re{An}), we employ only the  supermomenta variables $\eta^A$ and never their conjugates $\bar\eta_A$.

How can one construct purely holomorphic $SL(4|4)$ invariants from the momentum supertwistors? The idea  is suggested by the superfield analog \p{intr}\,%
\footnote{See also Refs.~\cite{ArkaniHamed:2009si,Hodges:2009hk} for a similar construction.}: Start with mixed chiral-antichiral invariants and integrate out the antichiral variables. Consider a set of holomorphic supertwistors $W_i$ (with $i=1,\ldots, n$), and
introduce a set of  auxiliary antiholomorphic ones $\widebar W^a$ (with $a=1,\ldots,k$).  Then the inner products $W_i\cdot \widebar W^a$, as well as any function of them $r(W_i\cdot \widebar W^a)$ will be automatically invariant under dual superconformal transformations. To get rid of the auxiliary supertwistors, it suffices to integrate  over $\widebar W^a$ with the $SL(4|4)$ invariant measure $ D^{4|4}  \widebar W{}^a =  d^4 \bar w^a\, d^4\bar \chi^a$,
\begin{align}\label{R1}
R_n^k(W)=\int \prod_{a=1}^k D^{4|4}  \widebar W{}^a\ r(W_i\cdot \widebar W{}^a)\,.
\end{align}
The function $r(W_i\cdot \widebar W{}^a)$ has degree $0$ in the Grassmann variables (counting $\chi$ and $\bar\chi$ as variables of opposite degree). Since the Grassmann integration in \p{R1} is equivalent to differentiation with respect to $\bar\chi$, it is clear that $R_n^k(W)$ has degree $4k$ in the Grassmann variables $\chi$ and, therefore,
it corresponds to an N${}^k$MHV amplitude.

Let us replace the function $r(W_i\cdot \widebar W{}^a)$ in \p{R1} by its Fourier integral
\begin{align}\label{F}
 r(W_i\cdot \widebar W{}^a) = (2\pi)^{-4k}
  \int Dt\,\tilde r(t)
 \exp\lr{i\sum_{a,i} t_a^i W_i\cdot \widebar W{}^a}\,,
\end{align}
where the integration measure $Dt$ will be discussed below.
Then, integration over $\widebar W{}^a$ yields a product of delta functions,
\begin{align}\label{Rk}
R_n^k(W)= \int Dt\,\tilde r(t) \prod_{a=1}^k \delta^{(4|4)}\left(\sum_{i=1}^n t_a^i W_i\right)\,,
\end{align}
where  $\tilde r(t)$ is some  function of $t_a^i$.
Notice that the dual superconformal invariant \p{msin} proposed by Mason and Skinner~\cite{Mason:2009qx} \footnote{Obviously, the same argument applies to the original proposal of Arkani-Hamed {\it et al.} \cite{ArkaniHamed:2009dn} for an integral representation of conventional superconformal invariants built from supertwistors.} has precisely this form, with the function $\tilde r(t)$ in \re{Rk} being part of the measure $\DT$ in  \p{msin}.

By construction, the integral \p{Rk} is invariant under dual superconformal $SL(4|4)$ transformations for {\em arbitrary}  $\tilde r(t)$. In particular, $R_n^k(W)$ is invariant under the global helicity transformations generated by the central charge ${\mathcal C}$,
Eq.\,\p{ducsusygen}.

In addition, the dual superconformal invariants we are seeking (suitable to appear in an amplitude) should also be invariant under the local helicity rescalings \p{W-h}. To ensure this property, the integration measure $Dt \,\tilde r(t)$  has to be invariant under $t_a^i \to  \zeta^{-1}_i t_a^i$. Further, we remark that the integrand $\prod_{a=1}^k \delta^{(4|4)}\left(\sum_{i=1}^n t_a^i W_i\right)$ in \re{Rk} is invariant under {\em local} $GL(k)$ transformations of the integration variables,
$t_a^i \to g_a{}^b(t) t_b^i$. This implies that the integration measure $\DT\equiv Dt\,\tilde r(t)$ should also have this local $GL(k)$ symmetry. Indeed, we can always integrate out the superfluous degrees of freedom present in $\DT$ but not in the integrand, thus reducing the measure to a locally $GL(k)$ invariant one. Such measures are discussed in detail in Sect.~\ref{pmdt}.

The relation \p{R1} can be considered as an integral transform of the non-chiral dual
superconformal invariant  $r(W_i\cdot \widebar W{}^a)$ into the chiral one $R_n^k(W)$.
The question arises whether this transform can be inverted.
Let us first use the local $GL(k)$ invariance in  \p{Rk} to fix a gauge, e.g.,
\begin{equation}\label{glkg}
    t^{\bi}_a= \delta^{\bi}_a\,, \quad (\text{for $\bi=1,\ldots,k$})\ ,
\end{equation}
after which \p{Rk} becomes
\begin{align}\label{Rk'}
R_n^k(W_a; W_\hi)= \int Dt\,\tilde r(t)\ \prod_{a=1}^k \delta^{(4|4)}\left(W_a + \sum_{\hi=k+1}^n t_a^i W_\hi\right)\,.
\end{align}
Next, let us perform a partial Fourier transform of $R_n^k(W_a; W_\hi)$ with respect to the supertwistors $W_a$ ($a=1,\ldots,k$), but  not to $W_\hi$ ($\hi=k+1,\ldots,n$):
\begin{equation}\label{paftr}
    \tilde R(\widebar W^a; W_\hi)= \int \prod_{a=1}^k D^{4|4}W_a\ \e^{-i\sum_a \bar W^a\cdot W_a} R_n^k(W_a; W_\hi) = \int Dt\,\tilde r(t^\hi)\,  \e^{i\sum_{a,\hi} t^\hi_a \widebar W^a\cdot W_\hi}\,.
\end{equation}
Comparing the right-hand side of this equation with \p{F}, we conclude that
\begin{equation}\label{mixed}
r(W_\hi\cdot\widebar W^a)= \tilde R(\widebar W^a; W_\hi)\,.
\end{equation}
This relation clearly shows a characteristic feature of the integral representation \p{Rk}: the invariant $ \tilde R(\widebar W^a; W_\hi)$, being a function of $n$ supertwistors, is obtained from another function $r(\tau^a_\hi)$ of $k(n-k)$ variables, by restricting it to the surface $\tau^a_\hi=W_\hi \cdot \widebar W^a$. In fact, what we see here is an example of an (inverse) John (or ``X-ray") transform.~\footnote{See Ref.~\cite{Gelfand} for a pedagogical introduction to the John integral transform and Ref.~\cite{Mason:2009sa}, Appendix B {for a review of its application in twistor theory.}}

\subsection{The general chiral momentum supertwistor invariants}\label{gcmsii}

So far we have constructed holomorphic (chiral) dual superconformal invariants as integrals of delta functions of linear combinations of holomorphic supertwistors, Eq.~\re{Rk}. But are we sure these are the most general invariants? We can give an affirmative answer to this question in two steps. First, in subsection \ref{sl11} we consider in detail the invariants of
the simplest superconformal symmetry $SL(1|1)$. Then, in subsection \ref{irci} we generalize to the case of interest, the invariants of $SL(4|4)$.

\subsubsection{Invariants of $SL(1|1)$} \label{sl11}

To simplify the problem of finding the most general superconformal invariant we consider $SL(1|1)$ instead of $SL(4|4)$. Take a function $R(w_i,\chi_i)$ of $n$ supertwistors $(w_i,\chi_i)$ and impose the invariance constraints
\begin{equation}\label{q-1}
    \mathcal{Q}\, R(w_i,\chi_i)=\tilde{\mathcal{Q}}\, R(w_i,\chi_i)= 0 \,.
\end{equation}
Here the supersymmetry generators
\begin{equation}\label{gensl11}
   \mathcal{Q}=  \sum_i w_i \frac{\pa}{\pa \chi_i}\,, \qquad  \tilde{\mathcal{Q}} = \sum_i \chi_i \frac{\pa}{\pa w_i}\,,
\end{equation}
satisfy the algebra
\begin{equation}\label{sl11alg}
    \{\mathcal{Q},\tilde{\mathcal{Q}}\}= \mathcal{C} \equiv \sum_i \left(w_i \frac{\pa}{\pa w_i} + \chi_i \frac{\pa}{\pa \chi_i}\right)\,,
\end{equation}
where $\mathcal{C} $ is the central charge or the total helicity (compare to \p{ducsusygen} and \p{ccal}). Then, the conditions \p{q-1} imply that the invariants have vanishing helicity,
$\mathcal{C}\, R=0$.

As before, we can use the transitive action of $\mathcal{Q}$ on the odd variables $\chi_i$ to fix the $\mathcal{Q}-$frame
\begin{equation}\label{q0}
    \chi_1=0\,.
\end{equation}
The dependence on the remaining $\chi_\hi$ $(\hi=2,\ldots,n)$ has the form
\begin{equation}\label{q1}
    R_n(w_1; w_\hi, \chi_\hi) = f(w) + \chi_\hi f^\hi(w) + \chi_{\hi_1} \chi_{\hi_2} f^{\hi_1\hi_2}(w) + \ldots + \chi_{\hi_1}\cdots\chi_{\hi_{n-1}} f^{\hi_1\cdots \hi_{n-1}}(w)\,,
\end{equation}
where $f^{\hi_1\ldots\hi_k}$ are functions of $w_1$ and $w_\hi$ with all the indices fully antisymmetrized. In this way we have imposed the first of the supersymmetry conditions \p{q-1}, $\mathcal{Q}\,R=0$.
Next, we turn to the second condition  $\tilde{\mathcal{Q}}\, R=0$. In the frame \p{q0} it becomes
\begin{equation}\label{q2}
     \tilde{\mathcal{Q}}\, R_n(w_1, w_\hi; \chi_\hi)=\sum_{\hi=2}^n \chi_\hi \frac{\pa}{\pa w_\hi} R_n=0\,. 
\end{equation}
Expanding in powers of $\chi$, we get  the conditions
\begin{align}\label{q02}
  \pa^\hi f(w) =\pa^{[\hi} f^{\hi_1]}(w)= \ldots= \pa^{[\hi} f^{\hi_1 \cdots \hi_{n-2}]}(w)= 0\,,
\end{align}
but we find {no restrictions} on $f^{\hi_1 \cdots \hi_{n-1}}(w)$.  Thus, in the Grassmann expansion \p{q1} the highest component $f^{\hi_1 \cdots \hi_{n-1}}(w)$ is {\it an arbitrary
function} of $w$, whereas all the other components satisfy the differential constraints \p{q02}.

The solution to \p{q02} has the following form
\begin{align}\label{pot}
f(w)=a_0(w_1)\,,\quad f^{\hi}(w) = \pa^\hi a(w)\,,\quad f^{\hi_1\hi_2} = \pa^{[\hi_1} a^{\hi_2]}(w)\,,\  \ldots \  f^{\hi_1 \cdots \hi_{n-2}} = \pa^{[\hi_1} a^{\hi_2 \cdots \hi_{n-2}]}(w)\,.
\end{align}
Since $a_0$ must have zero helicity and depends on $w_1$ only, it is reduced to a constant.
The ``potentials'' $a^{\hi_2 \cdots \hi_{k}}$, being arbitrary helicity neutral functions of $w$, are defined up to gauge transformations. Another way to see this is to write down the general solution to the constraint \p{q2} in the form $R_n = \tilde Q A(w,\chi)+\text{const}$, where the ``superpotential" $A$ is determined up to the gauge freedom $A\ \to \ A+ \tilde{\mathcal{Q}} \,\Lambda$.

Let us consider the $k-$th term in the expansion \p{q1} and let us try to write it down in the integral form analogous to \p{Rk}
  (for $1\leq k \leq n-1$):
\begin{equation}\label{q3}
     R_n^k =
    \chi_{\hi_1}\ldots\chi_{\hi_{k}} f^{\hi_1\cdots \hi_{k}}(w) =  \int Dt\ \tilde r_{k}(t) \prod_{a=1}^k \delta(t^i_a w_i)\delta(t^i_a \chi_i)\,,
\end{equation}
where the measure $Dt$ and the function $\tilde r_k(t)$ are supposed to have local $GL(k)$ invariance.
Expanding the Grassmann delta functions (in the frame \p{q0}) we can obtain explicit expressions for $f^{\hi_1\cdots \hi_{k}}(w)$:
\begin{eqnarray}
  k=1: && f^\hi(w) = \int Dt \ t^\hi_1\ \tilde r_1(t)\ \delta(t^1_1 w_1 + \sum_{\hi=2}^n t^\hi_1 w_\hi)\,,
  \nn\\
  k=2: && f^{\hi_1\hi_2}(w) = \int Dt \ \ep^{ab} t^{\hi_1}_a  t^{\hi_2}_b\ \tilde r_2(t)\ \prod_{a=1}^2 \delta(t^1_a w_1 + \sum_{\hi=2}^n t^\hi_a w_\hi)\,,\quad {\rm etc.}
  \label{q20}
\end{eqnarray}
Acting with $\pa/\pa w_\hi$ and antisymmetrizing the indices, it is easy to see that
the functions $f^{\hi_1\cdots \hi_{k}}(w)$ defined in this way indeed satisfy the constraints \p{q02}.
The question is if the integral transform \p{q20} (a version of the John transform, see the end of subsection \ref{hsti}) provides the most general solution to \p{q02}.

To answer this question we need to find a way to invert the  transform \p{q20}. The idea was already suggested in subsection \ref{hsti} and it relies on performing a partial Fourier transform of both sides of \p{q20}.
To illustrate the procedure, let us consider the simplest case $k=1$. Using the local $GL(1)$ invariance, we can fix the gauge $t_1^1=1$. Then we perform a partial Fourier transform of $ f^\hi(w)=\pa^\hi a(w)$ with respect to the variable $w_1$ and we replace $\tilde r(t^\hi_1)$ by its Fourier transform $r(\tau_\hi^1)$, to finally obtain (cf. \p{mixed})
\begin{equation}\label{q21}
    \tilde f^\hi(q^1; w_\hi) = \pa^\hi \tilde a(q^1; w_\hi)
    = i \left. \frac{\pa}{\pa \tau_\hi^1} r_1(\tau)\right\vert_{\tau_\hi^1 = q^1 w_\hi}
   \,,
\end{equation}
where the second relation is the expected most general solution to \p{q02}. We see that indeed we can always find a function $r_1(\tau_\hi^1)$, such that it reproduces an arbitrary $\tilde a(q^1; w_\hi)$. The next, less trivial example is the case $k=2$. Here we want to reproduce the closed two-form $f^{\hi\hj}(w)= \pa^\hi a^\hj(w)- \pa^\hj a^\hi(w)$. Note that the potential $a^\hi$ is defined up to the gauge freedom $a^\hi(w) \to a^\hi(w)+\pa^\hi\Lambda(w)$, which allows us to fix the gauge, e.g., $a^2(w)=0$. Further, we fix the $GL(2)$ gauge \p{glkg}, and repeating the steps above, we obtain the following expressions for the partial Fourier transform of  $f^{2\hi}(w)$ with respect to $w_1,w_2$:
\begin{equation}\label{q22}
    \tilde f^{2\hi}(q^1, q^2; w_\hi)=q^2 \tilde a^\hi(q^1, q^2; w_\hi) = i \left. \frac{\pa}{\pa \tau_\hi^1} r_2(\tau)\right\vert_{\tau_\hi^a = q^a w_\hi}\,, 
\end{equation}
(for $a=1,2$ and $\hi=3,\ldots,n$)
and similarly for $f^{\hi\hj}(w)$ with $\hi,\hj=3,\ldots,n$. Once again, we see that all the gauge-independent components of the potential $\tilde a^\hi(q^1, q^2; w_\hi)$ (with $\hi=3,\ldots,n$) are determined from the values of the derivatives of the function $r_2(\tau)$ on the surface $\tau_\hi^a = q^a w_\hi$.

These two examples illustrate that the integral transform \p{q20} does indeed provide the general solution to the supersymmetry constraints \p{q02}. We remark the redundancy in the transform -- to obtain the most general supersymmetry invariant $R_n^k$ it is sufficient to know its image $r_k(t)$ on a particular surface. This is typical for the John transform, unlike simpler integral transforms like Fourier or Radon \cite{Gelfand}.

Concluding this subsection, we would like to present an alternative interpretation of the result, suitable for generalization to the case of $SL(4|4)$. Consider again the expansion \p{q1}. As pointed out earlier, the highest component of this expansion involves the maximal number of Grassmann variables and plays a special r\^ole.
The reason for this is clear -- its $\tilde{\mathcal{Q}}-$variation is automatically suppressed in the frame \p{q0} and, therefore, the corresponding invariant $R_n^{k=n-1}$ is defined by an arbitrary function of $w_i$. How can we obtain invariants of lower Grassmann degree $1 \leq k \leq n-2$? The way is suggested by \p{q3} -- we need to restrict ourselves to a $k-$dimensional subspace of the $(n-1)-$dimensional space $(w_\hi, \chi_\hi)$, where we can still deal with the top term in the Grassmann expansion.

\subsubsection{Integral representation of the chiral invariants  of $SL(4|4)$}\label{irci}

Here we adapt the arguments developed above to the case of interest $SL(4|4)$. In principle, we should repeat each step, starting with the Grassmann expansion \p{q1}. The presence of $SL(4)$ R-symmetry indices of the $\chi$'s considerably complicates the expansion, without changing its nature. So, we prefer to skip this elaborate procedure and pass directly to the more intuitive argument mentioned at the end of the preceding subsection.

We start by using the 16 generators ${\mathcal Q}^{\hat A}_A$ from  \re{ducsusygen} to shift away four $\chi$'s in $R_n^k(W)$, e.g.,
\begin{align}\label{gauge1}
 \mbox{${\mathcal Q}-$frame:} \qqqquad \chi^{A}_{\bi}=0 \quad \text{ ($\bi=1,2,3,4$)}\,,
\end{align}
provided that the $4\times 4$ matrix $w^{A}_{\bi}$ built from the bosonic components of the
supertwistors
is invertible. {This regularity requirement excludes singular (contact term) superinvariants  like, for instance, $\delta^{(4)}(x_i-x_{j}) \delta^{(8)}(\q_i-\q_{j})$.%
\footnote{Note that the numerator in the MHV superamplitude \p{mhv}, if rewritten in dual superspace, becomes a dual superconformal invariant of this contact type \cite{Drummond:2008vq}.} It is easy to verify its invariance under \p{duinv}. However, for singular invariants of this type a frame like \p{gauge1} would make no sense: To shift away both $\q_i$ and $\q_{j}$ using the combined action of the $Q$ and $\bar S$ supersymmetry generators would require $x^2_{ij} \neq 0$, which is not the case.

The reason why we do not consider singular invariants is that the scattering amplitudes in $\cN=4$ SYM should be analytic functions of the Mandelstam variables with a complicated structure of physical poles and branch cuts controlled by on-shell unitarity~\cite{Bern:2007dw}.
This implies that the ratio function defined in \p{R} is given by regular (non-contact) superinvariants of the type \p{mbubl},\,%
\footnote{To be more precise, the invariant \p{mbubl} has pole singularities where any of the factor in the denominator vanishes. Some of these singularities are physical, that is they correspond to the expected analytic properties of the superamplitude, while other singularities are spurious and, therefore, should cancel against each other in the sum of superinvariants~\cite{Korchemsky:2009hm}. In the present paper we do not consider this issue and treat expressions of this type for generic kinematics, away from the poles.} so we can safely assume the validity of the frame \p{gauge1}.}

Now, we want to study the conditions for invariance of $R_n^k(W)$ under the other half of the supersymmetry generators  $\tilde {\mathcal Q}^{A}_{\hat A}$, in the fixed frame \p{gauge1}. There, the corresponding bosonic momentum twistors $w_{\bi}^A$
(with $\bi=1,2,3,4$) are invariant under $\tilde {\mathcal Q}$, whereas the remaining ones $w_{\hi}$  (with $\hi=5,\ldots,n$) are shifted by amounts proportional to $ \chi_\hi$.
Then, assuming that the invariant $R_n^k(W)$ is an unconstrained function $f(w_i)$ of all the bosonic twistor variables,
the only way to suppress its variation under  $\tilde {\mathcal Q}$ is to multiply it by the product of Grassmann delta function $\prod_{\hi=5}^n \delta^{(4)}(\chi_{\hi})$. As a result,
the corresponding dual superconformal invariant will have the maximal possible Grassmann degree $4k=4(n-4)$ and will take the following form in the frame \re{gauge1},
\begin{equation}\label{mhvdu}
 R^{(n-4)}_n(W) =   f(w_{\bi}, w_\hi) \prod_{\hi=5}^n \delta^{(4)}(\chi_{\hi}^A)\,.
\end{equation}
Here the bosonic momentum twistor variables $w_i$ are inert under all the supersymmetry generators, due to the fixed frame (for $w_{\bi}$) and  to the Grassmann delta functions (for $w_{\hi}$).

Working in the frame \p{gauge1}, it is not sufficient to demand invariance under $\tilde {\mathcal Q}$. We have to make sure that \p{mhvdu} is also annihilated by the anticommutator $\{{\mathcal Q}, \tilde {\mathcal Q}\}$, Eq.~\p{ccal}, otherwise the $\tilde {\mathcal Q}$ transformations will take us out of the frame \p{gauge1}. Let us start with the conformal (${\mathcal M}$) and R-symmetry (${\mathcal N}$) transformations. According to   \re{ducsusygen},
they rotate the $SL(4)$ indices of the bosonic and fermionic components of the supertwistors, $w_i^{\hat A}$ and $\chi_i^A$, respectively.
The Grassmann delta functions in the right-hand side of \re{mhvdu} are invariant under $SL(4)$ transformations of the $\chi$'s. The remaining bosonic factor, $f(w_i)$, has to be a function of {\it holomorphic} conformal invariants made of $w_i^{\hat A}$. These have the form of $4\times 4$ determinants
\begin{equation}\label{tau}
    u_{ijkl} =\frac1{4!} \ep_{{\hat A}{\hat B}{\hat C}{\hat D}}w^{\hat A}_{i}w^{\hat B}_{j}w^{\hat C}_{k}w^{\hat D}_{l} \equiv (w_i,w_j,w_k,w_l)\,,
\end{equation}
with some fixed values of the indices $i,j,k,l$. Then the invariant \p{mhvdu} takes the form
\begin{equation}\label{mhvdu''}
 R^{(n-4)}_n(W) =   f(u) \prod_{\hi=5}^n \delta^{(4)}(\chi_{\hi}) \,,
\end{equation}
where $u$ denotes the set of independent $SL(4)$ invariants \re{tau}.

Finally, we recall that $R^{(n-4)}_n(W)$ has to be invariant under local (point-dependent) helicity transformations \p{W-h}. As a result, the function $f(u)$ has to carry the helicity weights needed to compensate those of the Grassmann deltas in the right-hand side of \p{mhvdu''}. This can be achieved by including the factor
\begin{equation}\label{helfa}
    \prod_{\hi=5}^n \frac{(u_{1234})^3}{u_{\hi 234}u_{1\hi 34} u_{12\hi 4} u_{123\hi}}
\end{equation}
in  $f(u)$. The remaining freedom in $f(u)$ amounts to a function of the helicity neutral combinations of the conformal invariants \p{tau} (``cross-ratios"). The number of such (independent) cross-ratios is easily evaluated to be the total number $4n$ of all variables $w^{\hat A}_i$, minus the 15 parameters of the conformal group $SL(4)$, and minus the $n$ parameters of helicity rescalings at each point, which gives $3(n-5)$.\footnote{We recall that the central charge, or global helicity transformations should not be counted separately.}  Thus, the most general superinvariant corresponding to $k=n-4$ contains an arbitrary function of $3(n-5)$ helicity-less and conformally invariant variables.

The same general result \p{mhvdu} for superinvariants of maximal Grassmann degree can be reproduced starting from the integral formula \p{Rk}. Indeed, for $k=n-4$ the number of bosonic delta functions in \p{Rk} is $4(n-4)$, matching exactly the number of integration variables $t^i_a$, modulo the local $GL(k)$ invariance of the integral: $kn - k^2= 4(n-4)$. Then we can use these delta functions to express the $GL(k)$ gauge-independent part of the variables $t$ in terms of the $w$'s. Since the $t$'s are by definition conformal invariants, they will in fact be expressed in terms of the $u$'s. Finally, in the  frame \re{gauge1} and for $k=n-4$, the fermionic part of \p{Rk} reduces to that of \p{mhvdu''}, producing an extra bosonic factor $\lr{\det \|t_a^\hi \|}^4$. The important point in this argument is that the {\it arbitrary} function of conformal invariants $f(u)$ in \p{mhvdu''} matches the {\it arbitrary} function $\tilde r(t)$ in \p{Rk},
\begin{align}\label{intre}
 f(u) = \int Dt\,\tilde r(t) \lr{\det \|t_a^\hi \|}^4  \prod_{a=1}^{n-4} \delta^{(4)}(\sum_{i=1}^n t_a^i w_i)\,.
\end{align}
Indeed, in the right-hand side of \p{intre} we see the helicity-charged  factor $\lr{\det \|t_a^\hi \|}^4 $ which matches the factor \p{helfa} in the left-hand side. The function $\tilde r(t)$ has to contain a factor compensating the helicity of the differential form $Dt$ (see Sect.~\p{pmdt}). The remaining functional freedom in $\tilde r(t)$ amounts to a helicity neutral and conformally invariant function, therefore it effectively depends on the same cross-ratios as the function $f(u)$ in the left-hand side. In fact, the number of independent parameters $t^i_a$ is $kn$, minus the $k^2$ parameters of local $GL(k)$, minus the $n-1$ parameters of helicity rescalings.\footnote{The sum of all helicity parameters, or the central charge,  is identified with the $GL(k)$ weight.} This amounts to $(k-1)(n-k-1)$ gauge independent variables, which equals $3(n-5)$ for $k=n-4$, i.e. exactly the number of independent helicity-less conformal cross-ratios in $f(u)$.

We would like to point out that {in order} to achieve exact matching of
{the functional freedom of}
the superconformal invariants \p{Rk} and  \p{mhvdu''}, it is not {mandatory} to demand that $ R^{(n-4)}_n$ {has zero helicity with respect to each point. Indeed, even if we drop this condition, there still remains the requirement that the central charge vanishes (or equivalently  that the total helicity be equal to zero)}, as part of the superconformal symmetry $SL(4|4)$. Then, the number of variables in $f(u)$ will be $4n-16$, while in $\tilde r(t)$ we will have the same number of  free parameters, $kn - k^2= 4n-16$.

So far we have discussed the superinvariant of maximal Grassmann degree $4k=4(n-4)$ {corresponding to}  the $\overline{\rm MHV}$ amplitude.
Arriving at \re{mhvdu}, we have assumed that $f(w_i)$ is a generic function of all the bosonic twistor variables. This inevitably requires the presence of all the $(n-4)$ Grassmann delta functions in the right-hand side of \p{mhvdu}, needed to compensate the $\tilde {\mathcal Q}$ variations of the $w$'s. How can we relax this condition and obtain invariants of lower Grassmann degree $4k$, for any $k=1,\ldots, n-4$? Clearly, having fewer Grassmann delta functions, we will only be able to compensate the $\tilde {\mathcal Q}$ variations of a subset of the $w$'s. In other words, we need to restrict ourselves  to a $k-$dimensional subspace of the $(n-4)-$dimensional vector space spanned by the $w_\hi$'s, by considering the linear combinations $w_a(\tau)=\sum_{\hi=5}^n \tau_a^\hi w_\hi$. The arbitrary coefficients $\tau_a^\hi$ are assumed inert under the full superconformal algebra, but they have helicity weights, so that the combinations $w_a(\tau)$ have a ``collective" helicity weight $w_a(\tau)\ \to \ \zeta_a w_a(\tau)$. Their $\tilde {\mathcal Q}$ variations will be canceled by  the Grassmann delta functions depending on the same linear combinations of the $\chi$'s, $\chi_a(\tau)=\sum_{\hi=5}^n \tau_a^\hi \chi_\hi$. Then we can repeat the whole argument above, but this time with the starting point
\begin{equation}\label{mhvdu'}
 R^{k}_n(W) =   \int D\tau f(w_{\bi},w_a(\tau)) \prod_{a=1}^k \delta^{(4)}(\chi_a(\tau))
\end{equation}
replacing \p{mhvdu}. The reason why we integrate over the arbitrary auxiliary parameters $\tau$ is that they should not appear in the amplitude, which is a function of the supertwistors $W_i$ only.

Once again, the same result can be reproduced from \p{Rk}. This time we do not have enough bosonic delta functions to express all integration variables $t^\hi_a$ in terms of $w$'s. We can do this for only $4k$ of them, say  (see \cite{Mason:2009qx}):
\begin{align}\nn
& t_a^1 = -\frac{(w_a(t),w_2,w_3,w_4)}{(w_1,w_2,w_3,w_4)}\,,\qqqquad
t_a^2 = -\frac{(w_1,w_a(t),w_3,w_4)}{(w_1,w_2,w_3,w_4)}\,,\quad
\\
&
t_a^3 = -\frac{(w_1,w_2,w_a(t),w_4)}{(w_1,w_2,w_3,w_4)}\,,\qqqquad
t_a^4 = -\frac{(w_1,w_2,w_3,w_a(t))}{(w_1,w_2,w_3,w_4)}\,,
\end{align}
with $w_a(t) = \sum_{\hi=5}^n t_a^\hi w_\hi$. Notice that the dependence on
the $(n-4)$ twistors $w_\hi$ appears in $R_n^k$ only through the $k$ linear combinations $w_a(t)$. This allows us to establish the correspondence between the arbitrary conformally invariant and helicity-less function obtained from $f(w_{\bi},w_a(\tau))$ in \p{mhvdu'}, with the integral over $\tau$, and the analogous function $\tilde r(t)$ in  \p{Rk}, with the integral over $t$. Indeed, the function $f(w_{\bi},w_a)$ depends on $4(4+k)$ variables, minus 15 conformal parameters, and minus $(4+k)$ helicity parameters, which amounts to $3(k-1)$ independent variables. On the other hand, $\tilde r(t)$ has at least the same number of degrees of freedom,  $(k-1)(n-k-1) \geq 3(k-1)$, provided that $n \geq k+4$, which is indeed the case.

To summarize the discussion in subsection \ref{gcmsii}, we have naturally arrived at the most general form \p{msin} of the $n$-point dual superconformal invariant. Namely, $R_n^k$ is given by a product of $k$ graded delta functions depending on $k$ arbitrary linear combinations of $n$ momentum supertwistors and integrated over the space of parameters of the linear combinations. The number $k$ determines the Grassmann degree $4k$ of the invariant, which by definition corresponds to an N${}^k$MHV amplitude,
\begin{align}\label{1}
\mathcal{A}_n^{\rm N^kMHV} = \cA_n^{\rm MHV;0} R_n^k
= \cA_n^{\rm MHV;0} \int \DT\ \prod_{a=1}^k \delta^{(2)}(\sum_1^n  t_a^i\la_i) \delta^{(2)}(\sum_1^n  t_a^i x_i \ket{i}) \delta^{(4)}(\sum_1^n  t_a^i \q_i \ket{i})\,,
\end{align}
where $\cA_n^{\rm MHV;0}$ stands for the tree MHV superamplitude \p{mhv} and
we have restored the explicit expressions for the momentum supertwistor  components \p{ho}. This representation of the dual superconformal invariants leaves the freedom of choosing the integration measure $\DT = Dt\, \tilde r(t)$, discussed in the next subsection.

\subsection{Properties of the measure $\DT$}\label{pmdt}

We have to specify the integration measure $\DT=Dt\,\tilde r(t)$ in \p{Rk} and \re{1}. We recall that this measure should be invariant under the helicity rescalings
$t_a^i\to \zeta_i\,t_a^i$, as well as under local $GL(k)$ transformations  $t_a^i \to g_a{}^b(t) t_b^i$.

The invariance of the measure under local $GL(k)$ transformations effectively reduces the number of integration variables $t^i_a$ from $kn$ to $k(n-k)$. {As pointed out in  \cite{ArkaniHamed:2009dn}, this $k(n-k)-$dimensional space can be regarded as a Grassmannian $G(k,n)$, the space of $k-$dimensional subspaces in an $n-$dimensional vector space, or alternatively, the coset of $GL(n)$ divided by its parabolic subgroup:}
\begin{equation}\label{1'}
    G(k,n) = \frac{GL(n)}{GL(k) \times GL(n-k) \times N} \,,
\end{equation}
where $N$ is the subgroup of lower block-triangular matrices. A natural integration measure on $G(k,n)$ was proposed in \cite{Mason:2009qx}. It has the form
\begin{equation}\label{mogeme}
    \DT = \tilde r(t)  \,  {D^{k(n-k)}t} \,,
\end{equation}
with a particular weight function $\tilde r(t)$ to be specified below (see Eq.~\p{49}). The differential form in \p{mogeme} is defined as follows:
\begin{equation}\label{47}
   D^{k(n-k)}t = t_{i_{1,1} \ldots i_{1,n-k}} \ldots t_{i_{k,1} \ldots i_{k,n-k}} (d^kt)^{i_{1,1} \ldots i_{k,1}}\wedge \cdots \wedge (d^kt)^{i_{1,n-k} \ldots i_{k,n-k}}\,,
\end{equation}
where
\begin{align}\nn
  &  t_{i_1 \ldots i_{n-k}} = \frac{1}{n! k!} \ep_{i_1 \ldots i_n} \ep^{a_1 \ldots a_k} t_{a_1}^{i_{n-k+1}} \cdots t_{a_k}^{i_{n}}\,, \qquad
    \\ \label{48}
  &  (d^kt)^{i_{1} \ldots i_{k}} = \frac{1}{k!} \ep^{a_1 \ldots a_k} dt_{a_1}^{i_{1}}\wedge \cdots \wedge dt_{a_k}^{i_{k}}\,.
\end{align}
It is manifestly covariant under global $GL(n)$ transformations of the indices $i$ of $t^i_a$. In addition, it is covariant under local  $GL(k)$ transformations $t_a^i \ \to g_a^b(t) t^i_b$.  Indeed, when the differential  in each $dt^{i_l}_{a_l}$ acts on the $GL(k)$ matrix $g_a^b(t)$,  the variable $t^{i_l}_{a_l}$ gets free and it finds its match in the factors $t_{i_{1,1} \ldots i_{1,n-k}} \cdots t_{i_{k,1} \ldots i_{k,n-k}}$ to annihilate it because of antisymmetrization.
Thus, the measure \p{47} transforms with a certain local $GL(k)$  weight. Further, it has a total helicity weight $k$ at each point, since for every value of  the index $i$ the variable  $t^i_a$ occurs $k$ times in \p{47}.

To obtain an invariant measure $\DT$, the differential form $D^{k(n-k)}t$ has to be multiplied
by a weight factor  $\tilde r(t)$ which will compensate the $GL(k)$
and helicity weights of $D^{k(n-k)}t$. Since the variables $t^i_a$ transform under the fundamental representation of $GL(k)$ with respect to the index $a$, the only way to construct a $GL(k)$ covariant out of them is to form the minors of the rectangular matrix $t_a^i$ made from any $k$ columns labeled by $i_1, i_2, \ldots, i_k$:
\begin{equation}\label{mino}
    T_k^{(i_1, i_2, \ldots, i_k)} = \frac{1}{k!} \ep^{a_1 a_2 \ldots a_k} t_{a_1}^{i_1} t_{a_2}^{i_2} \cdots t_{a_k}^{i_k}\,.
\end{equation}
These minors have a $GL(k)$ weight, as well as the helicity weights of the points labeled by the $GL(n)$ indices $i_1, i_2, \ldots, i_k$. Notice that the global $GL(n)$ itself is inevitably broken by the minors.

Now, we can compensate the weights of the differential form \p{47} by taking the product of $n$  minors, such that each label $i=1,2,\ldots,n$ appears exactly $k$ times. One natural choice made in \cite{ArkaniHamed:2009dn,Mason:2009qx} is to consider the minors made from the {\it consecutive} columns $i, i+1,\ldots,i+k-1$:
\begin{equation}\label{50}
    T_k^{(i)} = \frac{1}{k!} \ep^{a_1a_2 \ldots a_k} t_{a_1}^i t_{a_2}^{i+1} \cdots t_{a_k}^{i+k-1}\,.
\end{equation}
With their help we can define the $GL(k)$ invariant and helicity neutral measure
\begin{equation}\label{49}
     \DT = \tilde r_0(t)  \,  {D^{k(n-k)}t}\,, \quad \mbox{with} \quad \tilde r_0(t) = \frac{1}{T_k^{(1)}T_k^{(2)}\ldots T_k^{(n)}}\,,
\end{equation}
where all indices in the denominator satisfy the periodicity condition, $i+n \equiv i$.

However, this choice is by no means unique. Since the variables $t$ are inert under dual superconformal symmetry, we have the freedom of multiplying the weight function $\tilde r_0(t)$ in \p{49} by an arbitrary function $\omega(t)$ with vanishing helicity and $GL(k)$ weight, \footnote{With the exception of the case $k=1$, where the only helicity invariant is the constant.}
\begin{equation}\label{freed}
    \tilde r(t) = \tilde r_0(t)\, \omega(t)\,.
\end{equation}
Such invariant functions can be built from the ``cross-ratios" of several minors of the general type \p{mino}.  For example, in the case $k=2$ the minors are $T_2^{(ij)} \equiv \ep^{ab} t^i_a t^j_b/2$, and the cross-ratios have the form
${T_2^{(ij)}T_2^{(kl)}}/(T_2^{(ik)}T_2^{(jl)})$ for any four different indices $i,j,k,l$.
Thus, the measure \p{49} can be modified by an arbitrary function of the cross-ratios
without affecting any of the symmetries discussed so far.

However, the situation changes when we impose the additional requirement of conventional conformal symmetry. We {shall argue below} that this extra symmetry rules out any invariant function $\omega(t)$ in \p{freed}, and so the weight factor $\tilde r(t)$ is uniquely fixed {to be of the form \p {49}}.  We do this in two steps. Firstly, we show that the twistor transform of the amplitude \p{1} with the special measure  \p{49} {is invariant under conventional superconformal  transformations}. Secondly, we prove that the only function $\omega(t)$, which is invariant under local $GL(k)$, helicity {\it and} conventional conformal symmetry, is the constant.

Conventional conformal invariance is a natural property of the tree-level amplitudes in ${\cal N}=4$ SYM theory. In the simplest case of the MHV amplitude \p{mhv}
this was shown by Witten in \cite{Witten:2003nn}. The difficulty stems from the non-local realization of conventional conformal symmetry in momentum space. This symmetry acts locally on points in the particle configuration space, but the transition to momentum space involves a non-local Fourier transform. As a result, the conformal generators are realized as {\it second-order} differential operators,
\begin{equation}\label{cogen}
    k_{\a\da} = \sum_{i=1}^n \frac{\pa^2}{\pa\la_i^{ \a} \tilde\la^{\da}_i}\,.
\end{equation}
The direct check of the symmetry of different types of amplitudes using these generators is certainly possible, although somewhat cumbersome.\footnote{
For the NMHV superamplitudes this was done in  \cite{Korchemsky:2009hm}. Recently, in \cite{Drummond:2010qh} the second-order generators were used for a direct proof of the conventional conformal invariance of the integral invariants \p{msin}. } However, for our second task, namely, proving that no conventional conformal invariants can be made out of the $t$'s, it is preferable to have a standard, local realization of the conventional conformal group. This can be achieved by first doing a  twistor transform~\cite{Witten:2003nn}, i.e. a half-Fourier integral with respect to $\tl$ and $\eta$, but not $\la$.\footnote{We point out that in the approach of Refs.~\cite{ArkaniHamed:2009dn,ArkaniHamed:2009vw} the twistor transform is done with respect to $\la$. The reason why we prefer the twistor transform with respect to $\tl$
is the manifest presence of $\la$'s in the MHV prefactor in \p{1}, which would complicate the task of Fourier transforming.   } It has the effect of making the conformal generators {\it first-order}, i.e. the conventional conformal group starts acting {\it locally}. The twistor transform of the amplitudes \p{1} is the subject of the next section.

\subsection{Regularization of the integral over the Grassmannian}\label{riog}

We conclude this section by a brief discussion of the regularization procedure for the integral over the Grassmannian manifold $\int \DT$. We wish to stress once again that our approach to the study of the conventional conformal properties of the dual superconformal invariants involves a twistor transform. This obliges us to deal with real {(bosonic)} twistor variables and Grassmannian parameters $t^i_a$. So, we have to give a meaning to the real integrals involving the singular measure \p{49}. {As we show below, this can be done and leads to the same results as the complex approach pursued in Refs.~\cite{ArkaniHamed:2009dn,Mason:2009qx}.}

As in illustration, let us consider a simple example of an integral $SL(1|1)$ invariant
$R_n^k(W)$ for  $n=3$, $k=1$ and with the measure \p{49}:
\begin{equation}\label{pv1}
    R_3^1(w,\chi) = \int_{-\infty}^\infty \frac{\ep_{ijk} t^i dt^j \wedge dt^k}{t^1 t^2 t^3}\ \delta(\sum_{i=1}^3 t^i w_i) \delta(\sum_{i=1}^3 t^i \chi_i)\,.
\end{equation}
Here all bosonic variables, $w_i$ and $t^i$, are real. We need to give a meaning of the integration over the $t$'s, in view of the pole singularities in the measure.

We start the evaluation of the integral by using the local $GL(1)$ freedom to fix a gauge, e.g., $t^1=1$, after which \p{pv1} becomes
\begin{equation}\label{pv2}
    R_3^1(w,\chi) = \int_{-\infty}^\infty \frac{dt^2 dt^3}{t^2 t^3}\ \delta(w_1+ t^2 w_2 + t^3 w_3) \delta(\chi_1+ t^2 \chi_2 + t^3 \chi_3)\,.
\end{equation}
We can use the bosonic delta function to carry out one of the integrations, e.g., with respect to $t^2$. The result is (modulo signs)
\begin{equation}\label{pv3}
    R_3^1(w,\chi) = \int_{-\infty}^\infty dt^3\  \frac{w_2\chi_1 - w_1 \chi_2 +  t^3 (w_2\chi_3 - w_3 \chi_2)}{w_2(w_1+t^3 w_3) t^3} \equiv \int_{-\infty}^\infty dt^3\  \frac{\beta +  t^3 \gamma}{(a+t^3) t^3}\,,
\end{equation}
where we have introduced a shorthand notation for the coefficients.

At this stage we realize that the remaining integral suffers from several problems. Firstly, it has poles on the real axis at $t^3=0$ and $t^3= - a$. Secondly, it is logarithmically divergent when $t^3 \to \pm\infty$. We can try to treat each of these singularities by the principal value prescription, i.e., by taking symmetric limits around each singular point:
\begin{equation}\label{pv4}
   {\rm PV} \int_{-\infty}^\infty dt^3\  \frac{\beta +  t^3 \gamma}{(a+t^3) t^3}
    = \lim_{A\to\infty, \ep\to0,\delta\to0}\left[\int_{-A}^{-a-\delta} + \int_{-a+\delta}^{-\ep} + \int_{\ep}^{A} \right] dt^3\  \frac{\beta +  t^3 \gamma}{(a+t^3) t^3} = 0\,.
\end{equation}
So, this regularization yields a trivial result.

Instead, we could replace one of the singular factors in the integrand by a delta function, e.g., $1/t^3 \ \to\ \delta(t^3)$, after which the integral in \p{pv3} becomes
\begin{equation}\label{pv5}
R_3^1(w,\chi) =    \int_{-\infty}^\infty dt^3\  \frac{\beta +  t^3 \gamma}{(a+t^3)}\ \delta(t^3) = \frac{w_2\chi_1 - w_1 \chi_2 }{w_1w_2}\,.
\end{equation}
We see that this kind of regularization solves all problems at once, leaving us with a well-defined and non-trivial result. An alternative choice would be to replace $1/(a+t^3) \ \to\ \delta(a+t^3)$, which leads to the result
\begin{equation}
R_3^1(w,\chi) =   -  \frac{w_3\chi_1 - w_3 \chi_2 }{w_1w_3}\,.
\end{equation}
instead of \p{pv5}.

{The approach of Refs.~\cite{ArkaniHamed:2009dn,Mason:2009qx} leads to the same expression for the invariants as the delta function regularization above. In it one starts by
writing down the  invariants \p{pv1} as multidimensional contour integrals in the complexified $t-$space and by making a particular choice of contour. For instance, for the
contour that encircles the pole at $t^3=0$ one applies the residue theorem  to reduce the $t^3-$integral \p{pv2} to
\begin{equation}\label{pv6}
    R_3^1(w,\chi) = \int\frac{dt^2}{t^2}\ \delta(w_1+ t^2 w_2)\ (\chi_1+ t^2 \chi_2 )\,,
\end{equation}
with a complex delta function $\delta(w_1+ t^2 w_2)$. The latter allows to do the remaining integral, arriving at the same result as in \p{pv5}. }

In summary, we have two ways to treat the singularities $1/t$ in the measure, as principal values or as delta functions. The former leads to a trivial result. The non-trivial prescription consists in first doing $4k$ integrals (in the case of $SL(4|4)$) with the help of the bosonic twistor deltas $\prod_{a=1}^k \delta^{(4)}(\sum_{i=1}^n t^i_a w_i)$, and then replacing all remaining pole singularities in the measure \p{49} by delta functions. We remark that $\delta(t)$ and $1/t$ have the same behavior under local $GL(k)$ (as usual, ignoring sign issues) and, as we shall see below, under the conventional conformal transformations of the $t$'s.

This prescription  can be also implemented by extending the
$t$ integrals into the complex plane. Indeed, after the integration
with the help of the $4k$ bosonic delta functions, we observe that the remaining
integrals involve a meromorphic
function of the $t$'s which admits a unique analytic continuation to the
complex $t-$plane. Then, the invariant can be rewritten as a
multi-dimensional contour integral and the choice of regularization
amounts to deforming the
contour around the poles at $T^{(i)}_k(t)=0$. The advantage of such a
representation is that the
invariants defined in this way are insensitive to the choice of the
space-time signature. In this form, the general expression for the superconformal invariants
coincides with the one proposed in Refs.~\cite{ArkaniHamed:2009dn,Mason:2009qx}.

\section{Twistor transform}\label{TT}

In this section we perform the  twistor transform of the amplitude  in \p{1} and prepare the ground for studying the conventional conformal properties. The twistor transform we are going to study is a Fourier transform of \re{1} with respect to the  $\bl$'s  and $\eta$'s,
\begin{align}\label{T}
T[\mathcal{A}_n](\{\la,\mu,\psi\}) =  \int  \prod_1^n \frac{d^2 \bl_i}{(2\pi)^2}\, d^4\eta_i\,  \e^{i\sum_1^n (\mu_{i\da} \bl_i^{\da} + \psi_{i A}  \eta_i^A )}\mathcal{A}_n(\{\la,\bl,\eta\})\,,
\end{align}
so that  $\mu_{i \,\da}$ and $\psi_{i\,A}$ are the Fourier conjugates of $\bl_i^{\da}$ and $\eta_{i}^A$, respectively. We recall that we work in a space-time with split signature  $(++--)$, in which case the $\bl$'s are real spinors, independent of the $\la$'s.

The twistor transform \re{T} can be performed in two steps. We start by casting the (super)momentum conservation delta functions from the MHV prefactor \p{mhv} in Fourier form \cite{Witten:2003nn},
\begin{equation}\label{9}
 \cA_n^{\rm MHV;0} = (2\pi)^{-4}\bigg({\prod_1^n \vev{i\, i+1}}\bigg)^{-1}    \int d^4X d^8\Theta\ \exp\left[ i \sum_{i=1}^n \left(  \la^\a_i  X_{\a \da}\tl_i^{\da} + \la^\a_i \Theta_{A\, \a} \eta_i^{A}\right)\right]\,,
\end{equation}
thus introducing {two} new integration variables, a real four-vector $X_{\a\da}$ and a chiral anticommuting spinor $\Theta_{A\,\a}$.

The second preparative step before the twistor transform is to replace the product of delta functions entering \re{1} by their Fourier integrals. To this end, we make use of the identities
\begin{align}\nn
\sum_i t_a^i x_i\ket{i} & =  \sum_i t_a^i (x_i-x_1)\ket{i} =  -\sum_{j=1}^n |j]\vev{j \rho_a^j}\,,
\\ \label{idd}
\sum_i t_a^i \theta^A_i\ket{i} & =  \sum_i t_a^i (\theta_i-\theta_1)^A\ket{i} =  -\sum_{j=1}^n \eta_j^A\vev{j \rho_a^j}\,,
\end{align}
obtained by using the ``conservation law" condition $\sum_1^n  t_a^i\la_i=0$ imposed by the first set of delta functions in \p{1}. In \p{idd} we have used the definitions  \re{2} and have introduced the notation for the composite ($t-$dependent) spinors
\begin{equation}\label{13}
  \ket{\rho^{i}_{a}} \equiv \sum_{j=1}^{i-1}  t_a^j \ket{j} \quad \mbox{(with $i=2,\ldots,n$)}\,, \qquad \ket{\rho^{1}_{a}} \equiv \sum_{j=1}^{n}  t_a^j \ket{j} \,.
\end{equation}
Taking \re{idd} into account, we obtain
\begin{align}
(2\pi)^2 \delta^{(2)}(\sum_1^n  t_a^i x_i \ket{i}) \delta^{(4)}(\sum_1^n  t_a^i \q_i \ket{i}) = \int
  d^2\tilde\rho^a\, d^4\xi^a\, \exp \left[-i \sum_{j=1}^n \sum_{a=1}^k  \vev{j\, \rho_a^j} \lr{[\tilde\rho^a j]+\xi_A^a \eta_j^A}\right].
\end{align}
After this, the Fourier integrals in \re{T} may be performed readily and  the twistor transform of the superamplitude \p{1} is given by
\begin{eqnarray}
  T[{\cal A}_n^k] &=& \int \DT \int  d^4X d^8\Theta\ \int\prod_{a=1}^k ( d^2 \rt^a d^4\xi^a) \nn \\
  &\times& \prod_{a=1}^k \delta^{(2)}(\rho^{1}_{a})\ \prod_{i=1}^n {\vev{i\, i+1}}^{-1}{  \delta^{(2)}(\mu_i^{\da} + (X_i)^{\da \a} \la_{i\,\a})\ \delta^{(4)}(\psi_{i\,A} + (\Theta_{i\,A})^{\a} \la_{i\,\a}) }\,, \label{11}
\end{eqnarray}
where we have introduced the ``moduli space coordinates"\footnote{This terminology was first used in Ref.~\cite{Witten:2003nn} and then taken over in Ref.~\cite{Korchemsky:2009jv}.}
\begin{eqnarray} \label{12}
  &&  X_i =   X + |\rt^a] \bra{\rho^{i}_a}\,,\qqquad  \Theta_i =  \Theta + \xi^a  \bra{\rho^{i}_a}\,,
\end{eqnarray}
with $\rho^{i}_a$ given by \re{13} (the sum over $a$ is implied). From  \re{12} it follows that the coordinates $X_i$ satisfy the relations
\begin{align}\label{4.8}
X_{i+1}-X_{i} =   |\rt^a]\lr{\bra{\rho^{i+1}_a}-\bra{\rho^{i}_a}} = t_a^{i} |\rt^a]\bra{i}\,.
\end{align}
In other words, the points $X_{i+1}$ and $X_i$ are lightlike separated,
$X_{i,i+1}^2=0$.

We remark the close similarity between the dual superspace with coordinates $(x_i, \q_i^A)$ and the moduli (or ``particle") superspace with coordinates $(X_i, \Theta_{iA})$. In both cases, we have $n$ points with lightlike separations, $x_{i,i+1} =  |i]\bra{i} $ in dual space and $X_{i,i+1} = - t_a^{i} |\rt^a]\bra{i}$ in moduli space.
{These points can be interpreted as defining the $n$ vertices of two lightlike polygons in dual and moduli spaces.}
The two spaces share the same chiral spinors $\la_\a^i$, but have different antichiral ones, $\tl^i_\da$ and $\rt_\da^a t_a^i$, respectively.  Note that the cases $k=0$ and $k=1$ are special. In the former we do not have any antichiral spinors $\rt^a$, so the lightlike $n-$gon with vertices at $X_i$ (with $i=1,\ldots,n$) shrinks to a point. In the latter all antichiral spinors are collinear, which means that all $X_i$ are coplanar (see Ref.~\cite{Korchemsky:2009jv}).

Regarded as a function on the supertwistor space with coordinates $(\lambda,\mu,\psi)$, the twistor transform \re{11} is localized on the configurations defined by the following set of constraints on  $\la$, $\mu$ and $\psi$:
\begin{align}
 \sum_{i=1}^n t_a^i \la_i^\a &= 0\,, \label{15}
\\[0mm]
\mu_i^{\da} + (X_i)^{\da \a} \la_{i\,\a} &= 0\,, \label{17}
\\[4mm]
\psi_{i\, A} + (\Theta_{i\,A})^{ \a} \la_{i\,\a} &= 0\,. \label{17'}
\end{align}
As was explained in \cite{Witten:2003nn,Korchemsky:2009jv}, the relations \p{17} and \p{17'}
define $n$ lines in  twistor space parameterized by the line moduli $(X_i, \Theta_i)$. Each particle lies on a separate twistor line and the lines of two adjacent particles with labels $i$ and $i+1$ intersect \cite{Korchemsky:2009jv}. Thus, the twistor transform of the amplitude \p{1} has its support on a configuration of $n$ intersecting twistor lines.

\section{Conventional conformal properties}\label{ocp}

By construction, the amplitude \p{1} is a dual superconformal  covariant,\footnote{The integral in \p{1} is a dual superconformal invariant, whereas the MHV prefactor carries a dual conformal weight \cite{Drummond:2008vq}. } but its properties under conventional superconformal symmetry are not manifest, {due to the non-local action of the symmetry in momentum space.} The reason why we did the twistor transform in the previous section was to simplify these transformations   and, as a consequence,
to make the superconformal properties of \p{1} more transparent.

\subsection{The MHV case}\label{mhvc}

We start by recalling the proof of conventional conformal invariance of the twistor transform of the MHV amplitude.\footnote{Here we follow the presentation of Ref.~\cite{Korchemsky:2009jv}, rather than the original one of Ref.~\cite{Witten:2003nn}.}
The latter corresponds to the case $k=0$ of the general expression, Eqs.~\p{11} and \p{12},
and it is obtained by dropping the integration over $t$, $\rt$, $\xi$ and by identifying all $X_i \equiv X$ and $\Theta_i \equiv \Theta$:
\begin{equation}\label{18'}
T[{\cal A}_n^0] = \int  d^4X d^8\Theta  \ \prod_{i=1}^n\frac{  \delta^{(2)}(\mu_i^{\da} + X^{\da \a} \la_{i\a}) \delta^{(4)}(\psi_{A\, i} + \Theta_{A}^{\a} \la_{i\a}) }{\vev{i\, i+1}}\,.
\end{equation}
By construction, $T[{\cal A}_n^0]$ is a function of $n$ sets of twistor variables $(\lambda_i, \mu_{i}, \psi_{A\, i})$ (with $i=1,\ldots,n$) describing the external particles.

{ Conventional superconformal symmetry acts on  $(\lambda^\a, \mu_{ \da}, \psi_{A})$ in the same way as dual conformal symmetry acts on the momentum supertwistors 
(recall \p{ducsusygen}). Its generators are \footnote{We denote the conventional conformal generators by small letters to distinguish them from analogous generators of dual conformal symmetry.}:
\begin{eqnarray}
  && q^{A\, \a} = \la^\a \frac{\pa}{\pa\psi_A}\,, \qqquad \bar q_A^{\da} = \psi_A \frac{\pa}{\pa\mu_{\da}}\,, \qqquad  p^{\da\a} = \la^\a \frac{\pa}{\pa\mu_{\da}}\,, \nn \\
  &&  s_{A\, \a} = \psi_A \frac{\pa}{\pa\la^\a}\,, \qqquad \bar s^A_{\da} = \mu_{\da} \frac{\pa}{\pa\psi_A}\,, \qqquad  k_{\a\da} = \mu_{\da} \frac{\pa}{\pa\la^\a}\,, \label{ordcsusygen}
\end{eqnarray}
in addition to the Lorentz ($SL(2)\times SL(2)$), R-symmetry ($SL(4)$), dilatation and central charge generators. The invariance of the integral \p{18'} under the Poincar\'e supersymmetry part of \p{ordcsusygen}, $q, \bar q$ and  $p$, is manifest, provided we accompany the transformations of the external variables by a suitable compensating transformations of the internal integration variables:
\begin{equation}\label{composu}
    q^{A\, \a} \Theta_{B\, \b} = \delta^A_B \delta^\a_\b\,, \qquad \bar q_A^{\da} X_{\b\db} = \delta^{\da}_{\db}  \Theta_{A\, \b}\,, \qquad p_{\a\da} X^{\db\b} = \delta^\b_\a\delta^{\db}_{\da} \,.
\end{equation}
Not surprisingly, they have the standard form of the chiral realization of Poincar\'e supersymmetry in the moduli superspace (compare with the analogous transformations of the dual superspace coordinates in \p{posusy}).

As explained in Sect.~\ref{dssms}, to extend Poincar\'e supersymmetry to the full superconformal algebra \p{ordcsusygen}, it is sufficient to prove invariance under conformal inversion $I$. In twistor space, like in momentum twistor space (recall \p{18ho}), inversion acts simply by exchanging $\la_\a$ with $\mu^\da$, while $\psi_A$ remains inert \cite{Korchemsky:2009jv},\footnote{In Ref.~\cite{Korchemsky:2009jv} we used a slightly different convention for the inversion of $\la$ and $\mu$, which led to some extra minus signs.}
\begin{equation}\label{18}
    I: \qquad (\la_\a)' = \mu^{\da}\,, \qquad (\mu^{\da})' = \la_\a \,, \qquad (\psi_A)' = \psi_A\,.
\end{equation}
As before, to verify the invariance of the integral \p{18'} under inversion, we have to accompany the transformation \p{18} of the {\it external variables} $(\la_i,\mu_i,\psi_i)$ by a compensating transformation of the integration variables $X$ and $\Theta$.
Once again, it takes the standard form of  inversion in chiral superspace, }
\begin{equation}\label{19}
    X' = X^{-1}\,, \qquad \Theta'= \Theta X^{-1} \,.
\end{equation}
In this way the twistor line equations \p{17} and \p{17'} are transformed covariantly,
\begin{equation}\label{20}
    (\mu_i + X\la_{i})' = X^{-1}(\mu_i + X\la_{i})\,, \qquad  (\psi_i + \Theta\la_{i})' = \psi_i + \Theta\la_{i}\,.
\end{equation}
So, each bosonic delta function  in \p{18'} produces a factor $X^2$, \footnote{Strictly speaking, the factor is $|X^2|$. As discussed in \cite{ArkaniHamed:2009dn,Mason:2009sa,Korchemsky:2009jv}, such sign factors cause the breakdown of global conformal invariance of the twistor transform. Here and it what follows we shall ignore this effect. We may say that we use conformal inversion as a convenient way to keep track of the transformations of the integrand and of the integration measure. For a proof we may switch to infinitesimal conformal transformations, which are not affected by such global effects.}
 while the fermionic delta functions stay invariant.
Next, taking into account the delta functions in \p{18'} we can use the twistor line equations \p{17}  to recast the transformation \p{18}  of the $\la$'s in the form
\begin{equation}\label{19'}
    (\la_{i\a})' = - X^{\da \b} \la_{i\b} \qquad \Longrightarrow \qquad \vev{i\, i+1}' = X^2\vev{i\, i+1}\,.
\end{equation}
So, the entire product $\prod_{i=1}^n $ in \p{18'} turns out {to be} invariant. In addition, we see from \p{19} that the measure is also invariant, which proves the invariance of the twistor transform \p{18'} under inversion, and thus its full conventional superconformal invariance.

\subsection{The N${}^k$MHV case}\label{nmhvc}

In the general, N${}^k$MHV case (for $k>0$) the twistor transform \p{11} involves
new integration variables, the commuting antichiral spinors $\rt^a$ and the Grassmann variables $\xi^a$, as well as {the parameters} $t^i_a$.
In addition, the twistor line equations \p{17} and \p{17'} are parameterized by the composite moduli $X_i$ and $\Theta_i$  defined in \p{12}.

\subsubsection{Transforming the twistor line equations}\label{ttle}

To verify the conformal invariance of the twistor transform \p{11}, we need to find out how all th integration variables therein should transform under inversion \p{18}, so that they can compensate the transformation of the external twistor variables. As before, the suitable
compensating transformations can be deduced from the requirement for the twistor line equations \p{17} and \p{17'} to be covariant under inversion. The first thought which comes to one's mind is to assume that the $X_i$ transform as points in moduli space,
\begin{equation}\label{naive}
    X_i' = X^{-1}_i\,,
\end{equation}
just like $X$ in \p{19}. This, together with \p{18}, clearly makes the twistor line equations \p{17} covariant,
\begin{equation}\label{22naive}
    (\mu_i + X_i\la_{i})' = X^{-1}_i( \mu_i  + X_i\la_{i})\,.
\end{equation}
 However, this choice leads to a system of linear equations for $(t_a^i)'$ which does not have
a solution for arbitrary $\mu_i$ and $\lambda_i$ (see Appendix~\ref{f1} for the detailed explanation).  The correct starting point, somewhat surprisingly,   turns out to be
\begin{equation}\label{22}
    (\mu_i + X_i\la_{i})' = X^{-1}( \mu_i  + X_i\la_{i})\,.
\end{equation}
{We have to stress the difference between the last two equations ($X_i^{-1}$ is replaced by $X^{-1}$) }
i.e., the twistor line equations must transform exactly as in the case $k=0$. For $i=1$, this makes sense since $X_1$ can be identified with $X$ in virtue of \p{15}. For $i\ge 2$ we notice that, unlike $X$, the moduli $X_i$ are composite objects \p{12} depending on the integration variables $\tilde\rho^a$ and $t_a^i$. So, our strategy is to use the definition \p{22} to first find  the correct transformation of $X_i$, and then derive from it the transformations of $\tilde\rho^a$ and $t_a^i$.

Taken alone, equation \re{22} is not sufficient to determine $X_i'$.
Another relation follows from the definition \p{12},
\begin{align}\label{22'}
(X_{i+1}-X_i)\ket{i}=0 \quad \Longrightarrow \quad (X_{i+1}'-X_i')|\mu_{i}]=0\,,
\end{align}
where the second relation is obtained by performing an inversion on the first and making use of \p{18}. Combining \p{22} and \p{22'}, we  obtain
\begin{align}\label{X-inv}
X_i' = X^{-1} + X^{-1} (X_i-X) \frac{\ket{i-1} [\mu_i| - \ket{i}[\mu_{i-1}| }{[\mu_i \mu_{i-1}] }\,,
\end{align}
where we used the standard notation for contraction of spinors, $[\mu_i \mu_{i-1}] \equiv
\mu_{i\,\da}  \mu_{i-1}^{\da}$.
It is important to realize that we derived this transformation rule {\it without using the twistor line equations} \p{17}. So, the rule \p{X-inv} holds for arbitrary $\mu_i$ and $\la_i$, not necessarily satisfying \re{17}. However, if we would make use of the twistor line equation
$\mu_i= -X_i\lambda_i$, relation \p{X-inv}  simplifies:
\begin{align}\label{X-on}
\lr{ X_i'}_{\rm on-shell}  {=} X^{-1} - X^{-1} (X_i-X)X_i^{-1} =  X_i^{-1}\,.
\end{align}
We notice that this is precisely the ``naive" transformation \p{naive}, but we have added the subscript ``on-shell" to indicate that this relation only holds on the shell of the twistor line equations \p{17}. {By ``on the shell of'' we mean the following}. Since the twistor transform \p{11} is localized on these lines,
we are allowed to apply \p{X-on} inside the integral \p{11}. This is not the case, however, of the integration measure  with respect to the variables   $t, \tilde\rho,\xi$ in \p{11}. To correctly compute the transformation  of this measure under inversion, it
is essential that we employ the ``off-shell" transformations of the parameters, i.e. those obtained without using the twistor line equations \re{17} and \re{17'}.

Now, let us apply \re{12} and substitute $X_i=X+\tilde\rho^a \rho_a^i$ into both sides of \p{X-inv},
\begin{align}\label{extract}
( \tilde\rho^a)' (\rho_a^i)' = X^{-1} \ket{\tilde\rho^a} \bra{\rho_a^i} \frac{\ket{i-1} [\mu_i| - \ket{i}[\mu_{i-1}| }{[\mu_i \mu_{i-1}] }\,.
\end{align}
From here we obtain the off-shell transformations of  $\tilde\rho^a$ and $\rho_a^i$,
\begin{align}\label{30}
 ( \tilde\rho^a)' = X^{-1} \ket{\tilde\rho^a} \,,\qquad
 (\rho_a^i)' =  \bra{\rho_a^i} \frac{\ket{i-1} [\mu_i| - \ket{i}[\mu_{i-1}| }{[\mu_i \mu_{i-1}] }\,,
\end{align}
with $\rho_a^i$ being the composite spinor defined in \p{12}. We recall that we are looking for a transformation under inversion, which must square to the identity. Indeed, repeating the inversion \p{30} twice, we immediately see that it satisfies this requirement.

We would like to stress again that the relations \p{30} are valid for arbitrary $\mu_i$ and $\lambda_i$ ``off shell". If we apply the twistor line equations \p{17},   the first relation in \p{30} does not change while the second one simplifies to
\begin{align}\label{30-on}
( \tilde\rho^a)'_{\rm on-shell} = X^{-1} \tilde\rho^a \,,\qquad
 (\rho_a^i)'_{\rm on-shell}=- \rho_a^iX_i^{-1}\,.
\end{align}

Let us now examine the transformation of the fermionic variables. For the fermionic
line equation \p{17'}, covariance is achieved if
\begin{equation}\label{22'''}
    (\psi_i + \Theta_i\la_{i})' = \psi_i + \Theta_i\la_{i}\,.
\end{equation}
This relation can be considered as the counterpart of the bosonic condition \re{22}. The only difference is that the fermionic line equation does not acquire a weight. As before, we supplement
\re{22'} with the relation $(\Theta_{i+1}-\Theta_i)\ket{i} =0$ following from \re{12} to get
\begin{align}\label{22''}
 (\Theta_{i+1}'-\Theta_i')|\mu_{i}] =0\,.
\end{align}
Combining \p{22'''} and \p{22''}, we obtain
\begin{align}\label{Theta-inv}
 \Theta_i' =  \Theta_i \frac{\ket{i}[\mu_{i-1}| -\ket{i-1} [\mu_i|}{[\mu_i \mu_{i-1}] }\,.
\end{align}
We recall that in the bosonic sector we had to distinguish two different forms
of the transformation of bosonic parameters (off-shell and on-shell) depending
on whether the bosonic twistor line equations were taken into account. The same happens in the fermionic sector. Namely, we could apply the fermionic twistor line equation $\psi_i+\Theta_i\lambda_i=0$ to eliminate $\Theta_i$ from the right-hand
side of \re{Theta-inv} and to obtain the on-shell version of the transformation \re{Theta-inv},
\begin{align}\label{Theta-on}
 (\Theta_i')_{\rm on-shell} =  \frac{\psi_{i}[\mu_{i-1}| -\psi_{i-1} [\mu_i|}{[\mu_i \mu_{i-1}] }\,.
\end{align}
 Notice however that in both versions (on-shell and off-shell) of the
transformation of $\Theta_i$ it is legitimate to apply the bosonic twistor line equation
$\mu_i+X_i\lambda_i=0$.
In this way, \re{Theta-inv} can be further simplified to
\begin{align}\label{Th-inv}
 \Theta_i' =  \Theta_i X_i^{-1}\,.
\end{align}
Once again, we recover the ``naive" transformation of $\Theta_i$ as of a superspace coordinate, similar to  \p{19}.

As follows from the definition \re{12}, $\Theta_i$  is expressed in terms of the fermionic moduli $\Theta$, the composite spinors $\rho_a^i$ and fermionic variables $\xi^a$. We already know the transformations of $\Theta$, Eq.~\re{19}, and of $\rho_a^i$, Eq.~\re{30}. In close analogy with $\rt^a$ in the bosonic case, we can use \re{Theta-inv} to obtain the transformation properties of $\xi^a$. Substituting $\Theta_i = \Theta+\xi^a \rho_a^i$
into \re{Theta-inv} and taking into account  \re{19} and \re{30}, we get (upon using the bosonic line equations)
\begin{equation}\label{29'}
    (\xi^a)'= \xi^a + \bra{\Theta}X^{-1}|\rt^a]\,.
\end{equation}
Unlike all previous cases, this transformation is inhomogeneous, $\xi$ is shifted by a composite fermion. But this is not a problem, we only need \p{29'} when discussing the transformation of the measure $\prod_{a=1}^k d^4\xi^a$, which clearly stays invariant under such shifts.

\subsubsection{Transforming the variables $t^i_a$}\label{ttt}

To prove the invariance of the twistor transform \p{11} under inversion, we still need
to find the transformation properties of the variables $t^i_a$. This can be done
by examining the  inversion property \re{30} of  the composite spinor $\rho_a^i$ in its explicit form in terms of $t^i_a$ and $\la_i$, Eq.~\re{13}.

Let us define the new scalar variables
\begin{align}\label{C-t}
 C_a^i=\vev{\rho_a^{i} i} = \sum_{j=1}^{i-1} t_a^j \vev{j i}\,.
\end{align}
A characteristic feature of these variables is that they are invariant under inversion.\footnote{The conformally invariant variables $C_a^i$ are closely related to those used in \cite{ArkaniHamed:2009dn}.}
Indeed, it follows from \re{18} and \re{C-t} that
\begin{align}\label{inv}
 (C_a^i)'= \vev{\rho_a^i \, i}' = [( \rho_a^i)' \mu_i] = C_a^i\,,
\end{align}
where in the last relation we applied \p{30}.

Taking into account the identity $\vev{\rho_a^{i+1} i} = \vev{\rho_a^{i} i}$ which
follows from the definition \re{13}, we can invert the relation \re{C-t} as follows,
\begin{align}\label{rr1}
\ket{\rho_a^{i}}=\sum_{j=1}^{i-1} t_a^j \ket{j} =\frac{C_a^{i-1}\ket{i}-C_a^{i} \ket{i-1}}{\vev{i\, i-1}}\,.
\end{align}
We can use the first of these relations to express $t_a^i$ in terms of $\rho_a^i$,
\begin{align}
t_a^i  = \frac{\vev{i+1\,\rho_a^{i+1}}-\vev{i+1\,\rho_a^{i}}}{\vev{i+1\,i}}\,,
\end{align}
and then apply the second relation in \re{rr1} to get
\begin{align}   \label{t-C}
t_a^i   =  \frac{C_a^i \vev{i+1\, i-1} + C_a^{i-1} \vev{i\, i+1} + C_a^{i+1}\vev{i-1\, i}}{\vev{i-1\, i}\vev{i\, i+1}}\,.
\end{align}
Performing inversion on both sides of this relation and taking into account \re{inv}, we get
 \begin{align} \label{29}
(t_a^i)'  &
  =  \frac{C_a^i [\mu_{i+1}\mu_{i-1}] + C_a^{i-1} [\mu_{i}\mu_{i+1}]+ C_a^{i+1}[\mu_{i-1}\mu_{i}]}{[\mu_{i-1}\mu_{i}][\mu_{i}\mu_{i+1}]}\,.
\end{align}
It is easy to see that this transformation has the inversion property $I^2=\mathbb{I}$.

Replacing $C_a^i$ in \re{29} by its expression in terms of  $t_a^i$, Eq.~\re{C-t}, we find that \p{29} has the form of a $GL(n)$ transformation of the $t$'s,
\begin{equation}\label{32}
    (t_a^i)' = \sum_{j=1}^n t^j_a\, g^i_j\,,
\end{equation}
with $g^i_j$ being a {\it lower-triangular} matrix
\begin{equation}\label{33}
    g^i_j = \frac{\vev{j\, i}[\mu_{i+1}\mu_{i-1}]+ \vev{j\, i+1} [\mu_{i-1}\mu_{i}] + \vev{j\, i-1} [\mu_{i}\mu_{i+1}]}{[\mu_{i-1}\mu_{i}][\mu_{i}\mu_{i+1}]}\ \theta(j<i) + \frac{\vev{i\, i+1}}{[\mu_{i}\mu_{i+1}]}\ \delta^i_j\,.
\end{equation}
This property will be useful in studying the transformation of the integration measure \p{49}.

When we use the transformations \p{32} outside the differential form $D^{k(n-k)}t$ in the measure $\DT$, Eq.\,\p{49}, we can take advantage of the twistor line equations \p{17} to simplify the expression  in \p{33}. We replace $\ket{i} = - X^{-1}_i|\mu_i]$ in the first term and then apply the cyclic identity for the $\mu$'s. The result is
\begin{align}\label{40}
( g^i_j )_{\rm on-shell}= -\frac{\bra{j}X_{i}|\rt^b] t^i_b}{X^2_{i} X^2_{i+1}}\theta(j<i) + \frac{1}{X_{i+1}^2}\ \delta^i_j\,,
\end{align}
and relation \p{32} becomes
\begin{equation}\label{41}
    (t^i_a)'_{\rm on-shell} = \frac{1}{X^2_{i+1}} (\delta^b_a- \bra{\rho^{i}_a}X^{-1}_{i}|\rt^b]) t^i_b\,.
\end{equation}
As follows from their definitions \p{13} and \p{12}, the variables $\rho^{i}_a$  and $X_i$ entering this relation depend on the variables $t$. As a result,  relation \p{41}
defines a {\it nonlinear} transformation of $t_a^i$. The explicit form of this transformation
can be found in Appendix~\ref{tofm}.

\subsubsection{Conformal invariance of the twistor transform}\label{poci}

Given the inversion rules above, we easily see that all factors in the twistor transform \p{11}, except the measure $\DT$, transform homogeneously with conformal weights listed below:
\begin{align}\notag
& \int d^4X d^8\Theta  \Rightarrow  1\,,\qquad &&
\int \prod_{a=1}^k ( d^2 \rt^a d^4\xi^a)  \Rightarrow  (X^2)^{-k} \,,
&
\\[2mm] \notag
&
 \prod_{i=1}^n   \delta^{(2)}(\mu_{i} + X_i\ket{i})   \Rightarrow  (X^{2})^n
 \,,\qquad &&
 \prod_{i=1}^n  \delta^{(4)}(\psi_i + \Theta_i\ket{i})  \Rightarrow 1\,,
 &
\\[2mm] \label{34}
&
\prod_{a=1}^k \delta^{(2)}(\rho^{1}_a) \Rightarrow (X^{2})^k \,,\qquad
&&
\prod_{i=1}^n \vev{i\, i+1}^{-1}  \Rightarrow  \prod_{i=1}^n (X^{2}_i)^{-1} \,.
 &
\end{align}
In obtaining the last relation we have used the bosonic delta functions to convert $\mu_i$ into $-X_i\ket{i}$,
\begin{equation}\label{35}
    \vev{i\, i+1}' = [\mu_i\, \mu_{i+1}] = \vev{i|X_i X_{i+1}|i+1}= \vev{i|X_{i+1} X_{i+1}|i+1} = X^2_{i+1} \vev{i\, i+1}\,,
\end{equation}
where $X_{i+1}\ket{i} = X_{i}\ket{i}$ follows from the definition \p{12}.

Collecting all weights from \p{34}, we find the subtotal weight $X^{2n}/\prod_{i=1}^n X^{2}_i$, which should be compensated by the integration measure $\DT$ \p{49}.
In other words, the conformal invariance of the twistor transform \p{11} requires
the following transformation property of the integration measure over the parameters $t$:
\begin{align}\label{37}
[{\mathcal D}t]_{n,k}'
= \left({ (X^{2})^{-n}\prod_{i=1}^n X^{2}_i}\right) \times \DT\,.
\end{align}
As explained in Sect.~\ref{pmdt} (see \p{mogeme}), the integration measure $\DT$ admits the following representation
\begin{align}\label{37'}
\DT =  \tilde r(t)\, {D^{k(n-k)}t} \,,
\end{align}
where $D^{k(n-k)}t$ is the differential form defined in \p{47}
and the weight function $\tilde r(t)$ is to be determined from its transformation properties.
We can easily work out the transformation of the differential form  $D^{k(n-k)}t$ by observing that  \re{32} has the form of a {\it global} (i.e., $t$-independent) $GL(n)$ transformation with the  $GL(n)$ matrix $g$ given by \p{33}. Since $D^{k(n-k)}t$ is $GL(n)$ covariant, it acquires the weight
\begin{align}\label{ggk}
D^{k(n-k)}t' =(\det g)^k \times D^{k(n-k)}t\,.
\end{align}
The matrix $g$ being lower-triangular, its determinant is given by the product of the diagonal terms in \p{33},
\begin{equation}\label{38}
    (\det g)^k = \prod_{i=1}^n \frac{\vev{i\, i+1}^k}{[\mu_i\mu_{i+1}]^k} = \prod_{i=1}^n (X_i^{2})^{-k}\,,
\end{equation}
where in the second relation we used the bosonic twistor line equation to replace
$[\mu_i\mu_{i+1}]=\vev{i|X_i X_{i+1}|i+1}=X_{i+1}^2 \vev{i\,i+1}$.
Combining Eqs.~\p{37}, \p{37'} and \p{38}, we find that the weight function
in the measure \p{37'} should compensate the subtotal conformal weight
\begin{align}\label{func}
\tilde r(t') =   \left({ (X^{2})^{-n}\prod_{i=1}^n (X^{2}_i)^{k+1}}\right) \times \tilde r(t)\,.
\end{align}
This relation can be considered as a functional equation for the weight function
$\tilde r(t)$. Since $\tilde r(t)$ appears in the twistor transform \p{11} accompanied by the bosonic delta functions,  we are allowed to use the on-shell version \re{41} of $t'$.

\subsection{Particular solution}

Thus, the twistor transform \p{11} will have both dual and conventional conformal invariance
provided that the weight function
$\tilde r(t)$ satisfies  \re{func}. Here we show that a particular solution of this equation is given by the weight function $\tilde r_0(t)$ from \p{49}, and in the next section we prove that this solution is in fact unique.

Let us first examine  \re{func} for $k=1$. In this case,
the transformation \p{41} greatly simplifies,
\begin{equation}\label{42}
    (t^i)' = \frac{1}{X^2_i X^2_{i+1}} (X^2_{i}- \bra{\rho^{i}}X_{i}|\rt]) t^i = \frac{X^2}{X^2_i X^2_{i+1}} \,t^i\,,
\end{equation}
where we have used \p{12} to replace  $X^2  = (X_i-\rho^{i}\rt)^2 = X_i^2 - \bra{\rho^{i}}X_i|\rt]$. Substituting \p{42} into \p{func} we verify that the function
\begin{align}\label{k=1}
\tilde r_{k=1}(t) =\frac1{ t^1 t^2\ldots t^n}
\end{align}
satisfies \p{func} for $k=1$.

For $k\ge 2$ we have to deal with the general transformation \p{41}. A crucial observation is that, in spite of the complicated form of \p{41}, there exist new `collective'
variables $T^{(i)}_{k}$, which transform covariantly. They are given by the minors $T_k^{(i)}$, Eq.~\p{50}, built from $k$ consecutive columns of the matrix $t_a^i$ (for $k=1$ we have $T^{(i)}_{1}=t^i$). As shown in Appendix \ref{tofm}, the minors $T^{(i)}_{k}$ transform homogeneously under inversion \p{41} with the following conformal weight,
\begin{equation}\label{43}
    (T^{(i)}_{k} )' = \frac{X^2}{X^2_i \cdots X^2_{i+k}} T^{(i)}_{k}\,,
\end{equation}
where the product of weights in the denominator is cyclic, $n+i \equiv i$.
We would like to emphasize that this is true only for minors made of $k$ {\it adjacent} columns of the matrix $t^i_a$. The general minors \p{mino} do not have this property.  Then, it is straightforward to check that
for general $k$ the function
\begin{align}\label{ansatz}
\tilde r_0(t) = \frac1{T^{(1)}_k T^{(2)}_k\ldots T^{(n)}_k}
\end{align}
satisfies the condition \p{func}.

This completes our proof that the twistor transform \p{11} with the measure \p{49} is invariant under conventional conformal symmetry, in addition to the dual one. In the next section we also show that this invariant is unique.

\section{Uniqueness of the amplitudes}\label{umdt}

The main question we address in this paper is whether the combination of dual and conventional conformal invariance completely fixes the form of the amplitude  \p{1}, and in particular, of the measure $\DT$ as given in \p{50} and \p{49}.

As discussed in Sect.~\ref{pmdt}, the measure could be modified by an arbitrary function $\omega(t)$ of the $t$'s  (see \p{freed}), which is invariant under local $GL(k)$ and helicity transformations. However, such a modification will immediately clash with the property of {conventional conformal symmetry} established in the previous section. We stress that dual conformal symmetry allows $\omega(t)$ to depend only on the inert integration variables $t$. Such a function is not affected by the twistor transform, so it will reappear in \p{11} as part of the modified measure $\DT$. On the other hand, we have already seen that the twistor transform \p{11} with the specific measure \p{49} and \p{50} does have conventional conformal symmetry. Hence, the function $\omega(t)$ we are trying to add must be an invariant of conventional conformal symmetry (in addition to local $GL(k)$ and helicity),   made of the variables $t$ alone.   We are now going to show that such invariants do not exist.

For the purposes of constructing invariants of conventional conformal symmetry it is preferable to switch from finite conformal transformations (inversions) to infinitesimal ones.  We derive their form in the next subsection. After that we formulate the corresponding Ward identities for the  function $\omega(t)$, and show that the only solution is a constant.

\subsection{Infinitesimal conformal transformations}\label{icb}

The conformal boosts (not necessarily infinitesimal) of $t_a^i$ are obtained by combining inversion \p{41} with a translation with parameter~\footnote{Here $k$ and $p$ denote finite conventional conformal and translation transformations. {We remind the reader that} $k$ and $p$ should not be confused with the dual generators $K$ and $P$. }
 $\kappa$, $k_\kappa=I p_\kappa I$.
According to \p{ordcsusygen}, the only twistor variables transforming under translation are $\mu_i$, with ${p}_\kappa \mu_i = - \kappa\la_{i}$. The invariance of the twistor line equation \p{17'} then implies ${p}_\kappa X_i = {p}_\kappa X = \kappa$, as it should be for moduli space coordinates. Using these relations, we get from \re{41}
\begin{align}
{p}_\kappa I [t_a^i] = \frac1{(X_{i+1}+\kappa)^2} \bigg\{\delta_a^b-\bra{\rho_a^{i}}({X_{i}+\kappa})^{-1}|\tilde\rho^b]\bigg\}  t_b^i\,.
\end{align}
Then we apply inversion to both sides of this equation and make use of \p{X-on}, \p{30-on} and \p{41} to obtain,
after some algebra,
\begin{align}
 {k}_\kappa  [ t_a^i ] = \frac1{X_{i+1}^2(X_{i+1}^{-1}+\kappa)^2}\bigg\{   \delta_a^b - \bra{\rho_a^{i}} (1+\kappa X_{i})^{-1}\kappa|\tilde\rho^b] \bigg\} t_b^i\,.
 \end{align}
 This relation defines a {\em finite} conformal  boost transformation of $t_a^i$ with an arbitrary parameter $\kappa$. Under infinitesimal transformations, $\kappa \to 0$, we have
${k}_\kappa  [ t_a^i ] =t_a^i+\delta_\kappa  t_a^i$ with
\begin{align}\notag
 \delta_\kappa  t_a^i &= - 2(\kappa\cdot X_{i+1}) t_a^i-\bra{\rho_a^{i}}\kappa |\tilde\rho^b]t_b^i\\
 &= - 2(\kappa\cdot X) t_a^i
-\sum_{j=1}^i \bra{j}\kappa |\tilde\rho^b]t_b^j t_a^i
 -\sum_{j=1}^{i-1} \bra{j}\kappa |\tilde\rho^b]t_a^jt_b^i\,. \label{60}
\end{align}
Here in the second relation we replaced $X_i$ and $\rho_a^{i}$ by their explicit
expressions \re{12} and \re{13}.

Finally, when studying the conformal properties of the integrand in \p{11}, we can use the invariance of the twistor transform under translations $X\to X+\kappa$ and $\mu_i\to \mu_i-\kappa\lambda_i$, to fix the {frame}
\begin{equation}\label{trga}
    \mbox{${p}-$frame:} \qquad X_{\a\da}=0\,.
\end{equation}
This frame is stable under infinitesimal conformal transformations, $\delta_\kappa X= -X \kappa X$, so we can apply \p{trga} in \p{60} to get
\begin{equation}\label{61}
    \delta_\kappa  t_a^i = -\sum_{j=1}^{i} \Omega_j^b(t_b^i t_a^j + t_b^j t_a^i)
+\Omega_i^bt_b^i t_a^i\,,
\end{equation}
where the notation was introduced for the transformation parameters
\begin{equation}\label{62}
   \Omega_j^b=\bra{j}\kappa |\tilde\rho^b]
\end{equation}
subject to the constraint (recall \p{15})
\begin{align}\label{101}
\sum_{j=1}^n \Omega_j^b\, t_a^j = 0\,.
\end{align}

\subsection{Uniqueness of the measure $\DT$}\label{uofm}

As explained in Sect.~\ref{pmdt}, dual superconformal symmetry leaves the freedom of modifying the integration measure $\DT$ with a function $\omega=\omega(t)$ of the variables $t_a^i$.


The function $w(t)$ has to satisfy three requirements. Firstly, it should be invariant under local $GL(k)$ transformations  $\delta t_a^i = g_a^b(t)\, t_b^i$.
Secondly, it must be invariant under
helicity rescalings $\delta t_a^i = \zeta_i \, t_a^i$. Finally, it must be invariant under the conformal transformations \p{61} with parameters $\Omega_j^b$ as defined in \p{62}. So, we are looking for an invariant function $w(t)$  satisfying
the condition
\begin{align}\label{100}
\delta \omega(t) = \sum_{a,i} \delta t_a^i \frac{\partial}{\partial t_a^i} \omega(t)=0\,,
\end{align}
with $\delta t_a^i$ given by the superposition of all three transformations,
\begin{align}\label{100'}
  \delta t_a^i = -\sum_{j=1}^{i} (t_b^i t_a^j + t_b^j t_a^i)\Omega_j^b + t_b^i t_a^i \Omega_i^b + g_a{}^b t_b^i + \zeta_i t_a^i\,.
\end{align}
Notice that when imposing conformal invariance we should treat the parameters
of the transformation as given not just by the four conformal boost parameters $\kappa^{\a\da}$, but by the much bigger, $k\times n$ matrix of parameters $\Omega_j^b$ defined in  \p{62}. The reason is that $\Omega_j^b$ also depends on the twistor variables $\lambda_j$ and on the integration variables $\rt^b$, while $\omega(t)$ is a function of $t^i_a$ only. Therefore, the $k n$ components of $\Omega_j^b$, modulo the $k^2$ conditions \p{101}, amount to $k(n-k)$ independent parameters.

The most convenient way of taking the local $GL(k)$ symmetry into account is to fix a gauge, in which the irrelevant degrees of freedom are eliminated from $t^i_a$. A natural gauge choice is obtained by splitting the index $1 \le i \le n$ into two subsets,  $\bi=1,\ldots,k$ and $\hi = k+1,\ldots,n$, and then setting
\begin{equation}\label{102}
    t_a^{\bi} = \delta_a^{\bi}\,,\qquad (\text{with $\bi=1,\ldots,k$})\,.
\end{equation}
The remaining true integration variables are then $t_a^{\hi}$.
Next, we have to make sure that the transformation \p{100'} does not take us out of the gauge \p{102}. This is achieved by imposing $k^2$ conditions $\delta  t_a^{\bi} =0$ and by using them to determine the parameters $g_a{}^b$ of the $GL(k)$ transformations. In this way, we obtain that $g_a{}^b$ is given by the upper triangular
matrix
\begin{equation}\label{104}
    g_a{}^b = \Omega^b_{a} \theta(b>a)+ \bigg(\sum_{c=1}^a \Omega^c_c
    - \zeta_a \bigg)\delta_a^b\,.
\end{equation}
Further, in the gauge \p{102} we can easily solve the constraint \p{101},
\begin{equation}\label{106}
    \Omega^b_a = -\sum_{\hj=k+1}^n\Omega^b_{\hj}\, t^{\hj}_a\,,
\end{equation}
and in what follows we can treat  $\Omega^b_{\hj}$ as independent parameters.
Substituting \re{104} and \re{106} into \re{100'} we obtain the transformation of the remaining variables $t_a^{\hi}$ (with $k+1\le \hi\le n$),
\begin{align}
    \delta  t_a^{\hi} =& (\zeta_\hi -\zeta_a)t_a^\hi-\sum_{\hj=k+1}^{\hi} (t_b^{\hi} t_a^{\hj} + t_b^{\hj} t_a^{\hi})\Omega^b_{\hj}
   + t_a^{\hi}t_b^{\hi}\Omega^b_{\hi}
    \nn\\
   & +\sum_{\hj=k+1}^n [t_b^{\hi} t_a^\hj \Omega_\hj^b\theta(b\le a)+t^\hi_a t^\hj_b  \Omega_\hj^b\theta(b\ge a+1)]\,, \label{107}
\end{align}
where the summation over $ b=1,\ldots,k$ is tacitly assumed.

Let us replace $\delta t_a^i$ in \re{100} by its explicit expression \re{107} (we recall that  $\delta  t_a^{\bi} =0$) and require that $\delta \omega(t)=0$ for
arbitrary parameters $\Omega_\hj^b$, $\zeta_\hi$ and $\zeta_a$.
The variation with respect to $\zeta_\hi$ and $\zeta_a$ yields the helicity conditions (with $1\le a \le k$ and $k+1\le \hi\le n$ fixed)
\begin{eqnarray}\label{hel}
  \sum_{\hj=k+1}^n t_a^\hj \frac{\partial \omega}{\partial t_a^\hj} =  \sum_{b=1}^k t_b^\hi \frac{\partial \omega}{\partial t_b^\hi} =0\,,\qquad \text{($a,\,\hi\ $ fixed)}
\end{eqnarray}
Taking these relations into account, the variation of $\delta \omega(t)$ with respect to $\Omega_\hj^b$ yields the conformal invariance condition
\begin{align} \label{EQ}
\sum_{\hi=k+1}^\hj \sum_{a=1}^k t_b^\hi  {t_a^\hj \frac{\partial \omega}{\partial t_a^\hi}}  =\sum_{\hi=k+1}^n\sum_{a=1}^b {t_b^\hi\frac{\partial \omega}{\partial t_a^\hi}}t_a^\hj \,,
\end{align}
{for arbitrary $k+1\le \hj\le n$ and $1\le b\le k$}.
The relations \re{hel} and \re{EQ} define the system of linear equations for ${\partial \omega}/{\partial t_a^\hi}$. In Appendix~\ref{app:conf} we show that its general solution is ${\partial \omega}/{\partial t_a^\hi}=0$ leading to
\begin{align}\label{trso}
\omega=\text{const} \,.
\end{align}
This proves our claim that the measure \p{49} is uniquely fixed by the combined dual and conventional conformal symmetries.

The following comments are in order.

The solution \p{trso} has been obtained under the tacit assumption that we were looking only for regular conformal invariants. Indeed, in deriving \p{trso} we have assumed that the $k-$dimensional minors of the matrix $t_a^\hi$ are different from zero. If we allow them to vanish, we will be dealing with singular, contact term solutions to the constraints \p{hel} and \p{EQ}. This issue has to do with the ambiguity in the definition of the singular measure \p{49} discussed in Sect.~\ref{riog}.

We have shown in Sect.~3.2.2 that, constructing the general form of the dual superconformal invariants, we can relax the condition of zero helicity at each
point and replace it by the weaker requirement for the total helicity to vanish.
We may ask how flexible the solution \re{trso} is, as far as the local helicity
condition is concerned. Examining \re{100} and \re{100'} we notice that this condition is encoded in the dependence of the parameters of the helicity transformations $\zeta_i$ on the particle number. To answer the above question,
we have to substitute $\zeta_1=\ldots=\zeta_n$ into \re{100'} and solve the resulting
equation \re{100} for $\omega(t)$. In this way, we find that the solution \re{trso}
is not unique anymore. In particular, in the simplest case of $k=1$, we show in Appendix~\ref{app:conf} that for an even number of particles
$n$, the general solution to the conformal symmetry constraints looks as
$\omega_{k=1}(t) = \varphi(t_1 t_3\ldots t_{n-1}/(t_2 t_4\ldots t_{n}))$ with an
arbitrary $\varphi(x)$. Thus, the solution \re{trso} heavily relies on the condition
of helicity neutrality {for each particle}.

\section{Conclusions}

The scattering amplitudes in planar $\cN=4$ SYM theory have dual
and conventional superconformal symmetries, exact at tree level and anomalous at loop level. In this paper, we found
the general form of the invariants of both symmetries. The main difficulty in
implementing these symmetries is due to the fact that they
cannot be simultaneously realized in a local way. In the standard on-shell superspace
formulation, with the
scattering amplitudes considered as functions of the external
(super)momenta organized into momentum supertwistors, the dual
conformal symmetry acts linearly while the generators of the conventional
conformal symmetry
are second-order differential operators. In this formulation, it
becomes straightforward
to construct the most general dual superconformal invariants in the form of an
integral over some auxiliary scalar $t-$parameters, Eq.~\p{Rk}. This integral
representation involves the weight function
$\tilde r(t)$ which is not fixed uniquely by the dual symmetry alone.

In order to impose the conformal
symmetry constraints, we performed a twistor (half-Fourier) transform of the
amplitudes.
This linearizes the action of the conventional conformal transformations and, at the same
time, elucidates the geometric meaning of the invariants. Namely, the
$n-$particle invariants are localized on
configurations in twistor space, consisting of $n$ intersecting lines parameterized by the
moduli $X_i$
and their Grassmann counterparts $\Theta_i^A$. As was observed in
Refs.~\cite{Korchemsky:2009jv,Bullimore:2009cb}, the points $X_i$  define the vertices of an $n-$gon in the moduli space with lightlike edges $X_{i,i+1}^2=0$.  Quite
remarkably, the same configuration naturally appears in the dual $x-$space,
with the only difference that the edges of the $n-$gon coincide with the massless particle momenta, $x_{i,i+1}=p_i$. {It should be pointed out, however, that the number of lines in the moduli space of the twistor transform is effectively smaller, as already observed in \cite{Korchemsky:2009jv} in the case of the N${}^k$MHV tree-level superamplitudes.  The reason is the necessity to replace some of the singular factors in the measure \p{49} by delta functions (or, equivalently, to choose contours encircling some of the $t$'s, see Sect.~\ref{riog}). This reduces the number of distinct points $X_i$ in moduli space (see \p{4.8}), and so some of the $n$ twistor lines coincide. \footnote{We thank David Skinner for a discussion on this point.} }

After the twistor transform, the conformal generators
are given by first-order differential operators. Examining their action on
the general dual superconformal invariant \p{Rk}, we found that the conventional conformal symmetry induces a nonlinear transformation on the $t$ parameters. Then, the requirement for \p{Rk} to be invariant with respect to both symmetries leads to a
set of differential equations for the weight function $\tilde r(t)$. We
demonstrated that these equations have the unique solution
\p{ansatz}, given by the product of minors made from the consecutive columns of the
matrix $t_a^i$. The resulting superconformal invariants
generalize the known tree-level and one-loop invariants~\cite{Drummond:2008vq,Drummond:2008bq,Drummond:2008cr} {and coincide with the recently proposed expression for the leading singularities of the scattering amplitudes in $\cN=4$ SYM
\cite{ArkaniHamed:2009dn,Mason:2009qx}}. As discussed in \cite{Bullimore:2009cb,Kaplan:2009mh,ArkaniHamed:2009dg}, they have an interesting interpretation as
multi-dimensional contour integrals over a Grassmannian.

{We would like to emphasize that when discussing the properties of the
integration measure \p{49} we have to make sure that  the integration in \p{Rk}
over the real $t_a^i$ is well defined. By virtue of the local $GL(k)$ invariance, this
integral is $k(n-k)-$dimensional. In addition, the $4k$ bosonic delta functions
localize the $t-$integral
on hypersurfaces of co-dimension $k(n-k-4)$. A prescription how to make this remaining integral
well defined was described in Sect.~\ref{riog}. It is important to realize that the choice of prescription
and/or of the integration contour in the complex $t-$plane is not
dictated by the symmetries. Different choices of the
integration contour lead to different superconformal invariants \cite{ArkaniHamed:2009dn}. In
particular, as was shown in Ref.~\cite{Mason:2009qx}, in the special case of
$n-$particle NMHV invariants, the known one-loop
NMHV invariants \p{mbubl} can be obtained from the general formulas \p{Rk} and \p{k=1}
by taking the residues at $n-5$ poles located at $t_i=0$ with
$i\neq a,a-1,b-1,b,c$. For five points in general positions, $i\neq
a,b,c,d,e$, the  same formula produces the most general NMHV invariant.
Moreover, the conjecture was put forward in Ref.~\cite{ArkaniHamed:2009dn} that
the leading singularity contributions to the all-loop N${}^k$MHV amplitudes
in $\cN=4$ SYM theory are described by the formula \p{Rk} for various specific
choices of the integration contours. It was further conjectured in Ref.~\cite{Bullimore:2009cb}
that the form of these contours is determined by the so-called primitive leading singularities
at $3k$ loops and below.}

The natural question
arises whether the same
relation can be extended to the subleading singularities of the superamplitudes.
Addressing this question, it is preferable to examine the ratio
function \p{R} rather than the amplitude itself. The reason for this is that the dual and
conventional conformal symmetries of the amplitudes are broken at loop level, while the
ratio function is expected to be dual conformal invariant at all loops~\cite{Drummond:2008vq}. This immediately implies
that the
ratio function should be a linear combination of dual conformal
(but not necessarily superconformal) invariants. At one loop, these
invariants are given
by the product of dual superconformal invariants and scalar coefficient
functions depending on the conformal cross-ratios of the dual (bosonic) $x-$variables~\cite{Drummond:2008vq,Drummond:2008bq,Brandhuber:2009xz,Elvang:2009ya}.
These functions are responsible for the breakdown of dual
supersymmetry. It
would be interesting to investigate whether the same pattern persists at
higher loops.%
\footnote{At higher loops, the computation of the ratio function becomes subtle in dimensional regularization due to the interference between $O(1/\epsilon)$ and $O(\epsilon)$ terms
and it may be advantageous to perform the analysis in the Coulomb branch of $\mathcal{N}=4$ SYM, in which case the infrared divergences are regulated by masses~\cite{Alday:2009zm}. }
If this is the case, then the all-loop superamplitudes will be given by linear
combinations of the invariants \p{Rk} multiplied by scalar coefficient
functions
which  depend on dual conformal cross-ratios and carry the dependence on
the coupling constant.

Finding the all-loop expressions for these scalar functions is a challenging problem.
We would like to emphasize that the conventional and dual superconformal symmetries alone are not powerful enough to completely determine the $\cN=4$ scattering amplitudes.
In Ref.~\cite{Korchemsky:2009hm} we used the example of the NMHV superamplitudes  to show that the combined action of both symmetries is  insufficient to fix all the freedom even at tree level. We argued that the additional information needed comes from the study of the analytic properties of the amplitudes. The requirement of absence of spurious singularities, together with the correct multi-particle singular behavior, determines the unique linear combination of superinvariants corresponding to the $n-$particle tree NMHV superamplitude.%
\footnote{Recently, it was shown in Ref.~\cite{Beisert:2010gn} that the same analyticity conditions can be implemented by modifying the symmetry generators in such a manner that the $\cN=4$ superamplitudes become invariant to one-loop order.
}
At loop level, the same requirement leads to nontrivial
constraints on the loop corrections to the scalar functions mentioned above. One possible way to determine these functions would be to extend the Wilson loop/MHV amplitude
duality to non-MHV amplitudes and to identify the dual object describing the
ratio function. We believe that the appearance of a lightlike $n-$gon in the moduli space of the twistor transform of the non-MHV superamplitudes is not accidental and that it will play an important role in the search for the dual object.

\medskip

\noindent
{\bf Note added.} A different approach to the problem discussed in this paper is presented in Ref.~\cite{Drummond:2010uq}.

\section*{Acknowledgments}

G.K. is grateful to Sergey Derkachov, Henrik Johansson and David Kosower for interesting
discussions and {for carefully reading the manuscript}. E.S. would like to thank Nima Arkani-Hamed, Freddy Cachazo, David Skinner and Edward Witten for useful discussions, and James Drummond and Livia Ferro for collaboration at the early stages of this work.  E.S. is grateful to the IAS-Princeton for hospitality during the final stage of this work. This work was supported
in part by the French Agence Nationale de la Recherche under grant
ANR-06-BLAN-0142, by the CNRS/RFFI grant 09-02-00308.

\appendix

\section{Appendix: Notation and conventions}\label{nc}

We work in four-dimensional space-time with split signature $(++--)$, in which the Lorentz group is $SL(2,\mathbb{R}) \times SL(2,\mathbb{R})$ and the conformal group is $SL(4,\mathbb{R})$. We use the standard two-component spinor notation for chiral spinors (e.g., $\la^\a$), antichiral spinors (e.g., $\mu^{\da}$) and four-vectors (e.g., $X^{\a\da}$ or $X_{\da\a}$). The indices are raised and lowered with the help of the Levi-Civita tensors:
\begin{align}
\la_{i\,\a} = \epsilon_{\a\b} \la^\b_i\,,\qquad \mu_{i\,\da} =\ep_{\da\db}\mu^\db_i
\end{align}
We also often use the bra-ket notation for contractions of spinor indices, e.g.,
\begin{equation}\label{a1}
    \vev{ij} = \la^\a_i \la_{j\a}\,, \qquad [\mu_i \mu_j] =  \mu_{i\,\da} \mu^{\da}_j\,, \qquad \bra{i}X|\rt] = [\rt|X\ket{i} = \la^\a_i  X_{\a\da}\rt^{\da}_j \,.
\end{equation}
Four-vectors are multiplied in matrix form as follows:
\begin{equation}\label{a2}
   X^{\db\a} X_{\a\da} = \delta_{\da}^{\db} X^2\,, \quad X_{\b\db} X^{\db\a} = \delta^\a_\b X^2\,, \quad  X_{\a\da} Y^{\da\a} = X^{\da\a} Y_{\a\da}= 2 X \cdot Y\,.
\end{equation}
The ``inverse" vector is defined by
\begin{equation}\label{a3}
    (X^{-1})^{\da\a} = \frac{X^{\da\a}}{X^2}\,, \qquad (X^{-1})^{\db\a} X_{\a\da}= \delta_{\da}^{\db}\,, \qquad X_{\b\db}  (X^{-1})^{\db\a}= \delta^\a_\b\,.
\end{equation}

\section{Appendix: On-shell versus off-shell transformations} \label{f1}

In Sect.\,\ref{ocp} we demonstrated that the conformal transformations of the different variables
in the twistor transform of the superamplitude \p{11} can be derived from the
condition that the twistor lines  \p{17} and \p{17'} transform covariantly.
We noticed that the corresponding conformal weight of the bosonic twistor line can
be chosen in two different forms,   Eqs.~\re{22naive} and \re{22}. Here we show that
the choice \re{22naive} is incompatible with the properties of the superamplitude \p{11}.

To begin with, we would like to emphasize that the variables $\mu_i$ and $\lambda_i$ are the bosonic twistor coordinates of the external particles. As such, they are independent from each other and their conformal properties are given by \p{18}. At the same time, the moduli $X_i$ are integrated over in \p{11} and we
have to use relations \re{22naive} and \re{22} to derive their transformation
properties under conformal inversion. We recall that $X_i$ are linear functions of the parameters $t_a^i$,
\begin{align}
X_i=X+|\tilde\rho^a]\bra{\rho_a^i} = X+\sum_{j=1}^{i-1}  t_a^j |\tilde\rho^a]  \bra{j}\,.
\end{align}
Substituting this relation into \p{22}, we first obtain the transformation of $\rho_a^i$, Eq.~\re{30}, and then use it to find the transformation of $t_a^i$, Eq.~\re{29}.
Had we used \re{22naive} instead of \re{22} as the  transformation of the twistor line equations, repeating the above analysis, we would find that $\rho_a^i$ transforms as
 \begin{align} \label{off}
 (\rho_a^i)'=- \rho_a^iX_i^{-1}\,.
\end{align}
This relation looks similar to  \re{30-on} but the important difference is that \re{off}
should hold off-shell, that is for arbitrary $\mu_i$ and $\lambda_i$, not necessarily
belonging to the twistor line \re{17}.
Replacing $\rho_a^i$ in \re{off} by its explicit expression \re{13} and taking into account \p{18}, we find the following system of linear
equations for $(t_a^i)'$:
\begin{align}\label{app:sys}
\sum_{j=1}^{i-1}  (t_a^j)' [\mu_j| = -\sum_{j=1}^{i-1}   t_a^j \bra{j} X_i^{-1}
\end{align}
(for $i=2,\ldots,n+1)$ and $a=1,\ldots,k$). Comparing the total number of equations,
$2kn$, with the number of unknown $(t_a^i)'$ we conclude that the system
is overcomplete  for arbitrary (off-shell) $\mu_i$ and $\lambda_i$. At the same time,
for $\mu_i$ and $\lambda_i$ belonging to the twistor line, $\mu_i+X_i\lambda_i=0$,
the system of equations \re{app:sys} has the solution \p{41}.

For general (off-shell) $\mu_i$ and $\la_i$ the transformation of $t_a^i$ is given by \p{32}. For $\mu_i$ and $\lambda_i$ satisfying the (on-shell) relation \re{17}, the same transformation simplifies to \re{41}.  The twistor transform \p{11} is localized on the
twistor line \re{17}.  This allows us to use the on-shell form of the transformation
inside the integral \re{11}. The question arises whether the same applies to the integration measure $\DT =  \tilde r(t) D^{k(n-k)}t$. In other words, whether the two forms (off-shell and on-shell) of the transformations of $t_a^i$ lead to the same transformation of the measure  $D^{k(n-k)}t$. To answer this question we consider the simplest case $k=1$.

For $k=1$ the integration measure takes the form
\begin{align}
D^{ (n-1)}t = \frac1{n!}\epsilon_{i_1\ldots i_{n}} t^{i_1}dt^{i_2}\wedge\ldots\wedge
dt^{i_n}
\end{align}
By construction, it is covariant under {\em global} $GL(n)$ transformations
\begin{align}\label{app:off}
(t^i)'= g^i_j \, t^j \,,\qquad D^{ (n-1)}t ' = \det g\cdot  D^{ (n-1)}t\,,
\end{align}
and under {\em local} $GL(1)$ transformations
\begin{align}\label{app:on}
(t^i)'= \varphi_i(t) \, t^i \,,\qquad D^{ (n-1)}t ' = \prod_{i=1}^n \varphi_i(t) \cdot  D^{ (n-1)}t\,.
\end{align}
The off-shell transformations of the parameters $t^i$ are defined in \re{32}. They
have the form of global $GL(n)$ transformations \re{app:off} with $g^i_j$ given by the lower-triangular matrix \re{33}. In this way, we find the conformal weight of the measure
under off-shell transformations as
\begin{align}\label{app:w1}
\det g =  \prod_{i=1}^n \frac{\vev{i\, i+1}^k}{[\mu_i\mu_{i+1}] } =   \frac1{X_1^{2}\ldots X_n^2}\,,
\end{align}
where in the second relation we used the bosonic twistor line equation to replace
$[\mu_i\mu_{i+1}]=\vev{i|X_i X_{i+1}|i+1}=X_{i+1}^2 \vev{i\,i+1}$.
The on-shell transformations of $t^i$ are given by \re{42}. Then, applying \re{app:on} for $\varphi_i(t) = X^2/(X_i^2 X_{i+1}^2)$ we find the conformal weight of the measure under on-shell transformations as
\begin{align}\label{app:w2}
\prod_{i=1}^n \varphi_i(t) = \frac{(X^2)^n}{(X_1^{2}\ldots X_n^2)^2}  \,.
\end{align}
We observe that the two weights \re{app:w1} and \re{app:w2} are different. This means that computing the conformal weight of the integration measure
$\DT$ we have to use the general (off-shell) form of the transformations of $t'$, valid for arbitrary $\mu_i$ and $\lambda_i$ not localized on the twistor line \re{17}.

\section{Appendix: Transformation of the minors}\label{tofm}

To prove Eq.~\p{43}, let us first rewrite \p{41} as follows:
\begin{equation}\label{44}
    (t^l_a)' =  \frac{\tau_a^l}{X^2_{l+1}}
\end{equation}
with $\tau_a^l$ being rectangular $k\times n$ matrix
\begin{equation}\label{45}
    \tau_a^l = \sum_{j=1}^{n} t^j_a\bigg(\delta_j^l - \bra{j}X^{-1}_{l}|\rt^b] t^l_b\ \theta(j<l)\bigg)\,.
\end{equation}
We want to compute $(T_k^{(i)})'$, i.e the determinant of the matrix $(t_{(i)})' \equiv \{(t^l_a)', \ i\le l\le  i+k-1\}$. This matrix differs from the matrix $\tau_{(i)} \equiv \{\tau^l_a, \ l=i, i+1, \ldots, i+k-1\}$ by the factors $1/X^{2}_{l+1}$  multiplying the $l$-th column of $\tau_{(i)}$. They can be pulled out of the determinant,
\begin{equation}\label{46}
     (T_k^{(i)})' = \frac{\det \tau_{(i)}}{X^2_{i+1} X^2_{i+2} \cdots X^2_{i+k}}\,.
\end{equation}
This accounts for most of the factors in the denominator in \p{43}. The remaining factor should come from the relation
\begin{equation}\label{51}
    \det \tau_{(i)} = \frac{X^2}{X^2_{i}} T_k^{(i)}\,.
\end{equation}
This is obtained with the help of the  identity (to be proved below)
\begin{align}\label{iden}
(\tau_{(i)})_a^l =(\Lambda_{(i)})_a^b\sum_{j=i}^{i+k-1}  t_b^j \, \Xi_{j}^l \,,\qquad
(l=i,\ldots,i+k-1)\,,
\end{align}
where
\begin{align}
(\Lambda_{(i)})_a^b = \delta_a^b- \bra{\rho^{i}_a} X_{i}^{-1} |\tilde\rho^b]\,,
\qqqquad
\Xi_{j}^l = \delta_{j}^l - \bra{j}X^{-1}_{l}|\rt^b] t^l_b\ \theta(j<l)
\end{align}
are $k\times k$ square matrices. In \p{iden} we see the product of three square matrices, hence
\begin{align}
 \det \tau_{(i)} =\det \Lambda_{(i)} \ T_k^{(i)}\ \det \Xi \,,
\end{align}
where $\det \Xi=1$ due to the triangular form of this matrix. Further,
\begin{align}
\det \Lambda_{(i)} = \exp\tr \log \Lambda_{(i)} = \exp\tr \log [1-X_{i}^{-1}(X_{i}-X)] =
\det (X_{i}^{-1} X) = \frac{X^2}{X_{i}^2}\,,
\end{align}
as follows from the property
\begin{equation}\label{52}
    (\Lambda_{(i)}\Lambda_{(i)})_a^c  = \delta_a^c - \bra{\rho^{i}_a} X_{i}^{-1} |\tilde\rho^b] \bra{\rho^{i}_b} X_{i}^{-1} |\tilde\rho^c] = \delta_a^c - \bra{\rho^{i}_a} X_{i}^{-1}(X_{i}-X)X_{i}^{-1} |\tilde\rho^c]\,.
\end{equation}
This completes the proof of \p{51}

Let us now prove \re{iden}. We start with \re{45} and split the sum over $j$ into two terms,
\begin{align}
\tau_a^l = t_a^l  - \sum_{j=i}^{l-1} \bra{j}X^{-1}_{l}|\rt^b] t_a^j t^l_b  - \sum_{j=1}^{i-1} \bra{j}X^{-1}_{l}|\rt^b] t_a^j t^l_b\,. \label{53}
\end{align}
Then we rewrite the second sum as
\begin{align}\notag
- \sum_{j=1}^{i-1} \bra{j}X^{-1}_{l}|\rt^b] t_a^j t^l_b
&= -
\sum_{j=1}^{i-1} \bra{j} X^{-1}_{i} |\rt^b] t_a^j t^l_b+\sum_{j=1}^{i-1} \bra{j}X^{-1}_{i}
(X_{l}-X_{i})X^{-1}_{l}|\rt^b] t_a^j t^l_b
\\
&=-
\sum_{j=1}^{i-1} \bra{j} X^{-1}_{i} |\rt^b] t_a^j t^l_b+\sum_{j=1}^{i-1} t_a^j \bra{j}X^{-1}_{i} |\tilde \rho^c] \sum_{m=i}^{l-1}t_c^m \bra{m}
 X^{-1}_{l}|\rt^b]  t^l_b \,.
\end{align}
Putting this back in \p{53}, we arrive at \re{iden}.

\section{Appendix: Solving the conformal symmetry constraints}\label{app:conf}

Let us rewrite \p{hel} and \p{EQ} in matrix form. Consider the rectangular matrices $t_a{}^\hi$ and $\omega_\hi{}^a \equiv \pa \omega/\pa t^\hi_a$, from which we can obtain two square matrices by left or right matrix multiplication:
\begin{align}\label{RL}
R_a{}^b \equiv t_a{}^\hi \omega_\hi{}^b\,,\qqqquad L_\hi{}^\hj \equiv \omega_\hi{}^a t_a{}^\hj\,.
\end{align}
These matrices have dimensions $k\times k$ and $(n-k)\times (n-k)$, respectively. From \p{hel} it follows  that both matrices have zeros on the main diagonal,
\begin{equation}\label{hel'}
    R_a{}^a =  L_\hi{}^\hi = 0 \quad \mbox{(no summation)}\,.
\end{equation}
Further, the relation \p{EQ} now reads (for $1\le b\le k$ and $k+1 \le \hj \le n$)
\begin{equation}\label{EQ'}
   \sum_{\hi=k+1}^\hj t_b{}^\hi L_\hi{}^\hj =\sum_{a=1}^b R_b{}^a t_a{}^\hj\,.
\end{equation}
It only involves the upper-triangular part of $L$ and the lower-triangular part of $R$. {In what follows we shall consider Eqs.~\re{hel'} and \re{EQ'} as a system of equations for the matrix $\omega_\hi{}^a$, and we shall demonstrate that their general solution is
\begin{align}\label{goal}
 \omega_\hi{}^a \equiv \pa \omega/\pa t^\hi_a = 0\,.
\end{align}
Note that in such an approach we need to divide by matrices made of the variables $t$. This implies certain non-singularity restrictions on the variables $t$, which will be discussed later.}

To begin with, let us examine  \re{hel'} and \re{EQ'} for different values of $\hj$ and $b$. Notice that in \re{EQ'} the indices $\hj$ and $b$ take $(n-k)$ and $k$ values, respectively.
Since relations \re{EQ'} are symmetric, we can assume without loss of generality that
$n-k\ge k$,  or equivalently $n\ge 2k$. For $\hj = k+1$ and $b=2$ we find that the left-hand side of \re{EQ'} vanishes in virtue of \re{hel'}, whereas the right-hand side of \re{EQ'} reduces to the single term $R_2{}^1 t_1{}^2$. Therefore,
assuming that $t_1{}^2\neq 0$, we find that $R_2{}^1=0$. In a similar manner, for $\hj=k+2$ and $b=1$ we obtain $L_{k+1}^{k+2}=0$, provided that $t_1^{k+1}\neq 0$. Continuing the analysis for
larger values of $\hj$ and $b$,  we can prove by induction that
\begin{align}\label{EQ1}
 L_\hi{}^{\hj} = R_b{}^a = 0\,,\qquad (1\le a\le b \le k, \quad   k+1\le \hi \le \hj \le 2k)\,,
\end{align}
provided that the square matrix $t_a{}^\hj$ (with $k+1\le\hj \le 2k$)  is invertible. In particular,
for $\hi=k+1$ we find from \re{EQ1} and \re{RL} that $\omega_{k+1}{}^a t_a{}^\hj=0$ for the $k$ different
values of the index $\hj$ listed in \re{EQ1}. This immediately implies that $\omega_{k+1}{}^a=0$,
in agreement with \re{goal}. Then, we use \re{RL} to deduce that $L_{k+1}{}^{\hj} =0$ for
arbitrary $\hj\le n$, thus extending the first relation in \re{EQ1} to $\hj>2k$. Moreover, substituting the second relation in \re{EQ1} into \re{EQ'} we find that both sides of \re{EQ'} vanish for arbitrary $\hj$,
\begin{align}\label{EQ2}
\sum_{\hi=k+1}^\hj t_b{}^\hi L_\hi{}^\hj = 0  \,,\qquad (\hj\le n)\,.
\end{align}
For $\hj\le 2k$ this is an identity, while for $\hj=2k+1$ we obtain a system of $k$ equations (for $b=1,\ldots,k$) whose solution is $L_\hi{}^{2k+1} = 0$ (with $i=k+1,\dots,2k+1$),
provided that the matrix $t_b{}^\hi$ is invertible. For $\hi=k+2$ we combine this relation
with the $(k-1)$ relations $L_{k+2}{}^{\hj}=0$, Eq.~\re{EQ1},  to deduce that $w_{k+2}{}^a t_a{}^{\hj}=0$.
The solution to this system of linear equations is $\omega_{k+2}{}^a=0$, in agreement with \re{goal}.
For $\hj\ge 2k+2$ the analysis of \re{EQ2} goes along the same lines and it yields \re{goal}
for $k+1\le i\le n$.
Thus, the general solution to \re{hel'} and \re{EQ'} is $\omega=\text{const}$.

Let us revisit the conformal Ward identity \re{100} in $k=1$ case,
but relaxing this time the condition for $\omega(t)$ to have zero
helicity at each point. For $\zeta=\zeta_1=\ldots=\zeta_n$,
infinitesimal conformal transformations \re{100'} of $t-$parameters look like
\begin{align}
\delta t^i = t^i\left[-2  \sum_{j=1}^{i}  t^j \Omega_j + t^i   \Omega_i
+  \zeta\right]\,,\qqqquad \sum_{i=1}^n \Omega_i t^i =0  \,,
\end{align}
where the second relation follows from \re{101}. Substituting $\delta t^i$ into  \re{100} we find that $\omega(t)$ has to satisfy the following conformal
Ward identities
 \begin{align}\label{app:W}
\left[ 2\sum_{i=j}^n t^i  \frac{\partial}{\partial t^i} - t^j \frac{\partial}{\partial t^j} \right] \omega(t) =\Lambda(t)\,,\qquad \sum_{i=1}^n t^i \frac{\partial \omega}{\partial t^i} = 0 \,,
\end{align}
for arbitrary $\Lambda(t)$. Here the second relation imposes the condition for the total helicity of $\omega(t)$ to vanish.
Solving the first relation in \re{app:W} for $j=n,n-1,\ldots$ we get
\begin{align}
 t^n\frac{\partial \omega}{\partial t^n}   =- t^{n-1}\frac{\partial \omega}{\partial t^{n-1}}   = \ldots = (-1)^{n-1}  t^1\frac{\partial  \omega}{\partial t^1}    =\Lambda(t)\,.
\end{align}
Substituting this expression into the second relation in \re{app:W} we find
that for odd number of particles $n$, the general solution is given by \re{trso},
whereas for even $n$ there exists a nontrivial solution
\begin{align}
w(t) = \varphi\lr{\frac{t_1 t_3 \ldots t_{n-1}}{ t_2 t_4 \ldots t_n}} \,,
\end{align}
with $\Lambda(x) = -x \,\varphi'(x)$.

\newpage


\end{document}